\documentclass[journal,transmag]{IEEEtran}
\ifCLASSINFOpdf
  \usepackage[pdftex]{graphicx}
  \DeclareGraphicsExtensions{.png}
\else
\fi

\usepackage[numbers]{natbib}
\usepackage{fnpct}

\usepackage{float}
\usepackage{multirow}
\usepackage{xcolor}
\usepackage{enumitem}

\begin{document}
\bstctlcite{IEEEexample:BSTcontrol}
%
\title{Cloud-Enabled High-Altitude Platform Systems: Challenges and Opportunities}


\author{Khaleel~Mershad,
        Hayssam~Dahrouj,
        Hadi~Sarieddeen,
        Basem~Shihada,
        Tareq~Al-Naffouri, \\
        and~Mohamed-Slim~Alouini
\thanks{Accepted for publication at the Frontiers in Communications and Networks,  Special Issue on Resource Allocation in Cloud-Radio Access Networks and Fog-Radio Access Networks for B5G Systems.}
\thanks{K. Mershad is with the Department of Computer and Communications Engineering, Arts, Sciences \& Technology University, Beirut, Lebanon.}
\thanks{H. Dahrouj is with the Center of Excellence for NEOM Research at King Abdullah University of Science and Technology (KAUST). (e-mail: hayssam.dahrouj@gmail.com)}
\thanks{H. Sarieddeen, B. Shihada, T. Al-Naffouri, and M.S. Alouini are with the Division of Computer, Electrical and Mathematical Sciences and Engineering, King Abdullah University of Science and Technology, Thuwal 23955-6900, Saudi Arabia. (e-mails: hadi.sarieddeen@kaust.edu.sa, basem.shihada@kaust.edu.sa, tareq.alnaffouri@kaust.edu.sa, slim.alouini@kaust.edu.sa).}
\thanks{Manuscript received May 28, 2021, accepted June 28, 2021.
Corresponding author: K. Mershad (email: khaleel.mershad@aul.edu.lb).}
}

\markboth{Journal of \LaTeX\ Class Files,~Vol.~XX, No.~X, Month~20XX}%
{Author \MakeLowercase{\textit{et al.}}: IEEE Transactions on }
%




\IEEEtitleabstractindextext{%
\begin{abstract}
Augmenting ground-level communications with flying networks, such as the high-altitude platform system (HAPS), is among the major innovative initiatives of the next generation of wireless systems (6G). Given HAPS quasi-static positioning at the stratosphere, HAPS-to-ground and HAPS-to-air connectivity frameworks are expected to be prolific in terms of data acquisition and computing, especially given the mild weather and quasi-constant wind speed characteristics of the stratospheric layer. This paper explores the opportunities stemming from the realization of cloud-enabled HAPS in the context of telecommunications applications and services. The paper first advocates for the potential physical advantages of deploying HAPS as flying data-centers, also known as super-macro base stations. The paper then describes various cloud services that can be offered from the HAPS and the merits that can be achieved by this integration, such as enhancing the quality, speed, and range of the offered services. The proposed services span a wide range of fields, including satellites, Internet of Things (IoT), ad hoc networks (such as sensor; vehicular; and aerial networks), gaming, and social networks. For each service, the paper illustrates the methods that would be used by cloud providers to offload the service data to the HAPS and enable the cloud customers to consume the service. The paper further sheds light on the challenges that need to be addressed for realizing practical cloud-enabled HAPS, mainly, those related to high energy, processing power, quality of service (QoS), and security considerations. Finally, the paper discusses some open issues on the topic, namely, HAPS mobility and message routing, HAPS security via blockchain and machine learning, artificial intelligence-based resource allocation in cloud-enabled HAPS, and integration with vertical heterogeneous networks.
\end{abstract}

}

\maketitle

\IEEEdisplaynontitleabstractindextext

%
\IEEEpeerreviewmaketitle

\section{Introduction}
\label{Sec:intro}
%
%
%
%
%
\IEEEPARstart{S}{everal} types of aerial networks are being considered for complementing the terrestrial communication infrastructures, which is emerging as one of the major innovative initiatives of the next generation of wireless systems (6G) \cite{saeed2020pointtopoint}. Such compound networks, also known as vertical heterogeneous networks (VHetNets) are expected to provide important services to ground systems, such as enhanced connectivity, backhauling, exploration and surveillance, rescue operations, traffic monitoring and control, etc. Some of these networks, such as the unmanned aerial vehicle (UAV) and tethered balloon (TB) networks, have already been used in several applications\footnote[1]{https://share-ng.sandia.gov/news/resources/news\_releases/solar\_balloons/} and projects \cite{knodler2019application} \cite{Alsamhi_drones} \cite{Alsamhi_B5G}. Other aerial networks, such as the High-Altitude Platform System (HAPS), have been extensively researched and are currently being considered by several projects for future deployment\footnote[2]{https://www.hapsmobile.com/en/}\footnote[3]{https://www.aircentre.org/projects/haps-high-altitude-platform-station/}. In this introduction section, we first illustrate the importance of utilizing aerial networks as part of future communication systems and the expected advantages. Next, we provide an overview of the basic elements of this paper, which are HAPS and cloud computing. Finally, we briefly describe the importance of integrating cloud computing into HAPS, which we detail in the remainder of the paper.

\subsection{Overview}
\label{Sec:overview}
The constant growth of interest in high-speed wireless communications causes the search for new solutions and new types of radio access networks. Although more than half of the global population is already connected to the Internet, there is still a need for higher broadband connectivity and telecommunication services in suburban, rural, and isolated areas that remain underserved. In addition, ground networks are susceptible to failures and degraded performance in many situations, such as natural disasters, severe weather conditions, and changing environments. For these reasons, several types of aerial networks are being considered as a means to provide both stable broadband connectivity to end-users and communication links between the mobile and core networks for backhauling purposes. These two types of applications could open the way to wireless broadband deployment in distant locations, such as mountains, seaside, and desert areas. The HAPS is expected to be among the major networks that will assist the ground infrastructure in future telecommunication systems. With its wide footprint (500 km radius per node) and reduced round-trip delay (0.13 to 0.33 milliseconds (ms)) \cite{kurt2020vision}, HAPS is ideal for low-latency and mobile applications that require continuous connectivity, as it relieves such applications from the burden of frequent handoffs that occur in ground networks.

In addition, as next-generation communication systems emerge, new high-data-rate applications become very prevalent. Consequently, network traffic has been growing so fast that current backhaul networks will soon fail to handle all traffic demands. A backhaul network provides connectivity between the cellular base stations (such as 4G eNBs and 5G gNBs) and the core network; it significantly impacts the performance of the whole network, and it is one of the major challenges in beyond 5G (B5G) and 6G. Towards supporting the backhauling requirements of future 6G networks, the development of a supplementary backhaul network is considered vital. As one of the solutions, a wireless backhaul network based on an aerial platform has been proposed by several research projects \cite{song2020analysis, challita2017network}. These backhaul networks could utilize one or more types of wireless signals (such as radio frequency (RF), millimeter waves (mmWaves), microwaves, lasers, and free space optics (FSO)) in order to provide self-sufficiency, flexibility, and encompass a wide range of application domains. In fact, Song \textit{et al.} \cite{song2020analysis} outlined several advantages of aerial-based wireless backhaul networks, namely, reduced cost, network scalability, easiness to deploy in any area, and guaranteed line-of-sight (LoS) propagation.

Another major advantage of HAPS is its ability to connect users in disconnected or weakly connected areas, and to provide to these users the various applications and services that are offered by public and private ground networks. One of the domains that will benefit from a strong HAPS constellation network is cloud computing. This is particularly the case since the HAPS would allow the cloud to reach a wider range of users, enhance the quality of service (QoS) of traditional cloud applications, and establish new cloud services that benefit from the unique characteristics of the HAPS, as we illustrate in this paper.

\subsection{Background}
\label{Sec:background}

\subsubsection{Cloud Computing}
\label{Sec:cloudcomputing}
Cloud computing (CC) is the process of supplying computing resources over the Internet. Instead of maintaining their own computing infrastructure or data centers, businesses can rent access to various services offered by cloud providers, such as digital applications, storage, databases, networks, and analytics tools. By utilizing these resources, companies can make use of any computational assets they require, and during the period they need them, without the need to acquire and sustain a physical, on-site infrastructure and framework. This new computing paradigm provides several advantages to customers, such as reduced costs, flexibility and efficiency, scalability, and information sharing.

Cloud computing is based on the virtualization model. Virtualization gives CC its unique characteristics, which are 1) on-demand self service, which allows each consumer to individually setup and manage computing resources such as server utilization and storage automatically without the need to contact the cloud provider, 2) broad network access, which regulates the access to facilities and resources that are available over the cloud network via standard procedures, 3) resource pooling, in which software and hardware computing resources are configured to meet the needs of multiple customers by dynamically allocating and reallocating each resource based on the customers’ needs, 4) rapid elasticity, which allows the provider to dynamically add or remove physical and virtual resources in such a way that they seem infinite to the customer, and 5) measured service, in which the service provider manages and optimizes the customers’ usage of resources by means of a resource measurement technique that is suitable to the type of the provided service.

In cloud computing, many types of resources and digital applications can be offered as a service (which led to the proliferation of the term XaaS, where X is the name of the provided service). Services may be in the form of hardware, software, storage, platform, infrastructure, database, and many others. The National Institute of Standards and Technology (NIST) presented three major categories of services, known as the CC service models, which are: 

\begin{enumerate}
\item
Infrastructure as a service (IaaS), which allows the user to rent computing capabilities, such as storage facilities, networks, processing power, and virtual private servers, on-demand and over the web. These resources are priced using a "pay as you go" method where the price is computed based on factors such as how much storage was required or the amount of processing power that was consumed over a specific period.
\item
Software as a service (SaaS), in which the cloud provider offers and manages a wide range of applications and software tools that the customers access and execute over the web. SaaS frees the customer from many responsibilities, such as the need for software maintenance, infrastructure administration, data availability, information security, and all the other concerns related to preserving the application in an active and operating state.
\item
Platform as a service (PaaS), which combines features from both SaaS and IaaS. In this service model, the cloud platform provides the customer the middleware, database management system, operating system, web server, and development tools. These tools constitute a remote environment where users can develop, build, and execute their software projects without the need to own any hardware or software.
\end{enumerate} 
 
\subsubsection{High-altitude Platform System}
\label{Sec:HAPSbackground}
A HAPS is a network of aerial stations that operate in the stratosphere at an altitude of around 17-20 km. Triggered by technological advances in the fields of communication systems, solar power systems, antenna arrays, and autonomous flight, and encouraged by the thriving industry environment, the HAPS has arisen as an essential constituent of future wireless networks \cite{kurt2020vision}. Owing to the distinctive characteristics of the stratosphere, a HAPS node can remain at a quasi-stationary position. In addition, the HAPS node will be equipped with a flight control system that enables the ground center to control its mobility and direct it towards a certain location whenever needed. Note that a HAPS node does not need a takeoff vehicle, i.e., it can move using its own power, and it can be fetched back to Earth, upgraded, and redeployed.

Previous research on HAPS focused on utilizing this technology in the areas of ubiquitous connectivity and disaster relief applications. Very recently, a new model was proposed in which the HAPS is considered to be a practical potential for future wireless communication networks \cite{kurt2020vision}. Following the developments in solar panel systems, HAPS nodes can utilize solar power that can be combined with storage batteries as the primary methods for energy supply; HAPS stations possess large surfaces that can house solar panel films. Moreover, because of its low-delay characteristics compared to evolving satellite networks, the HAPS can efficiently offer wireless services to the users of terrestrial networks.

The HAPS can have encouraging benefits over other wireless system infrastructures. According to \cite{kurt2020vision}, the low-altitude placement of HAPS, as compared to satellite networks, enables them to make use of better channel conditions and higher signal-to-noise ratio (SNR). In addition, the almost stationary position of HAPS nodes allows full utilization of their capabilities, preventing the existence of a significant Doppler shift and enabling a long airborne duration with marginal energy consumption. Moreover, due to its large volume, a HAPS node is appropriate for multiple-input multiple-output (MIMO) and massive-MIMO antenna arrangements. When assisted with multi-antenna arrays, the HAPS node can produce highly directional 3D beams with narrow beamwidths that increase the signal-to-interference-plus-noise ratio (SINR) for all users. Finally, due to its relatively low altitudes (as compared to other satellites networks), the HAPS end-to-end delay is estimated to be between 0.13 to 0.33 ms \cite{kurt2020vision} which is favorable for delay-sensitive applications. In general, with large antenna array deployments and low power requirements, the HAPS can support a wide range of fixed and mobile user terminals to satisfy various service demands \cite{aragon2008high}.

When deployed, HAPS will be the third main layer of communications infrastructure alongside terrestrial and satellite systems. HAPS networks promise to offer wide coverage and large capacity to serve highly populated suburban and rural areas, as well as isolated areas with weak connectivity, hence supplementing the current wired and wireless infrastructure. Arum \textit{et al.} \cite{arum2020review} explore the possibility of extending the achievable wireless coverage of terrestrial systems by utilizing HAPS. The authors show that efficient intelligent radio resource and topology management can mitigate inter-system interference and ensure the coexistence of terrestrial systems and HAPS with improved overall system performance. Hence, the ultimate goal of the HAPS network is to enhance the performance and accessibility of current communication systems. This is achieved by supporting a high capacity, comparable to that provided by terrestrial systems, and a wide coverage area, comparable to that provided by satellites. Therefore, HAPS networks are not projected to substitute state-of-the-art terrestrial and satellite technologies but rather to assist them in a collaborative and consolidated fashion.

\subsubsection{Cloud-enabled HAPS}
\label{Sec:cloudHAPS}
The previous subsection illustrates the critical role that the HAPS is expected to play in future communications systems and applications. With its position as a mediator between terrestrial and satellite networks and its large footprint that connects multiple terrestrial and low-altitude networks, the HAPS will become the communication hub connecting users to many networks and systems. This provides an opportunity for cloud providers to use the HAPS strategic position to reach a wider range of customers and improve the quality of their services and their infrastructure capabilities. As mentioned in \cite{kurt2020vision}, the HAPS station is expected to have a large size that enables it to operate as an aerial data center. In addition to data and computational offloading, which was pointed out in \cite{kurt2020vision}, the HAPS will enable cloud providers to deploy and offer their services from within the HAPS data centers. From the one side, the HAPS connects to users in rural and remote areas either directly or via low-altitude gateways, which enables users in these areas to get access to the cloud network via the HAPS. Examples of low-altitude gateways include low-altitude aerial base stations and tethered balloons (TBs). From the other side, the HAPS assists the cloud network in urban areas by providing additional resources and capabilities to improve the QoS of cloud services. 

\begin{table*}[ht]
\centering
\caption{ Comparison with sample recent overview papers on HAPS.} \label{tab:par}
\begin{tabular}{|p{2cm}|p{8cm}|p{7cm}|c|} 
    \hline
    \textbf{Reference}  & \textbf{Focus} &\textbf{Comparison with the current paper} \\
    \hline
     {Kurt \textit{et al.}\cite{kurt2020vision}}  &  {Comprehensively surveys the literature on state-of-the-art HAPS networks and provides a future vision of unrealized potentials and the corresponding challenges. The paper addresses HAPS design, resource allocation, and topology management criteria. }  & \multirow{12}{7cm}{This paper describes on a high level the main concepts of a novel future technology, as opposed to surveying an existing topic. To the best of our knowledge, the interplay between HAPS and cloud computing has never been studied before. We promote applications that make use of the HAPS-cloud integration and discuss their corresponding methods and challenges. The exhibition in this paper is one step forward towards establishing innovative initiatives for flying networks towards 6G.}  \\ \cline{1-2}
    {Arum \textit{et al.} \cite{arum2020review}}  & {Overviews HAPS prospects for service provisioning in remote areas and explores the corresponding techniques for intelligent radio resource allocation, topology management, and coverage extension.} & \\  \cline{1-2}
       {d’Oliveira \textit{et al.} \cite{d2016high}}  & {Surveys the HAPS technological trends at the time of writing from an aeronautical engineering perspective (not a communications perspective).} &\\ \cline{1-2}
    {Qiu \textit{et al.} \cite{Qiu8869712}}  & {Envisions a software-defined integrated network architecture that is cross-layer, comprising high- and low-altitude platforms alongside terrestrial cellular networks. The paper further motivates the prospects and challenges of the corresponding use cases, and proposes a proof-of-concept case study.} & \\
     \hline
\end{tabular}
\end{table*}

We finalize this section by illustrating in Table \ref{tab:par} the differences between this paper and previous survey/review papers on HAPS. Kurt \textit{et al.} \cite{kurt2020vision} describe a future vision of a powerful HAPS network, in addition to its applications, design requirements, and resource allocation methods. Arum \textit{et al.} \cite{arum2020review} explore the corresponding techniques for intelligent radio resource allocation, topology management, and coverage extension in HAPS. The authors of \cite{d2016high} provide a comprehensive review of the HAPS technological trends from an aeronautical engineering perspective. On the other hand, Qiu \textit{et al.} \cite{Qiu8869712} propose a cross-layer integrated network architecture that comprises multiple high- and low-altitude platforms alongside terrestrial cellular networks. For an extended list of books and survey papers on HAPS systems, the reader can check the references within \cite{kurt2020vision}. Unlike these papers, the essential purpose of our paper is not to provide a full survey or tutorial of HAPS infrastructure and applications. Rather, to the best of our knowledge, our paper is the first of its kind which overviews the various aspects related to the integration of cloud computing services into the HAPS. The HAPS-cloud platform that we propose in this paper is a one step forward towards establishing a framework for offering cloud services from HAPS data centers, which promises to be a vital area for a multitude of future research directions.

The paper next presents the opportunities that can be realized by integrating various cloud services within the HAPS and the corresponding cloud-type applications that would utilize the HAPS for enhancing the quality, range, and type of offered services. The paper then presents some of the challenges that need to be addressed for realizing practical cloud-enabled HAPS, mainly those related to the high energy, processing power, QoS, and security considerations. Finally, the paper sheds light on some open research directions in the field, namely, HAPS mobility and message routing, HAPS security via blockchain and machine learning (ML), artificial intelligence (AI)-based resource allocation in cloud-enabled HAPS, and integration with vertical heterogeneous networks.  
To the best of the authors' knowledge, this paper is the first of its kind to advocate for the use of cloud-enabled HAPS from a holistic perspective. The paper presents the challenges and open issues of such deployment, which promises to be an active area of research in the context of 6G systems development and cross-layer optimization.

\section{Opportunities}
\label{Sec:opportunities}

\subsection{Traditional CC Services}
\label{Sec:traditionalCC}
\subsubsection{Overview}
The number of user applications that utilize the cloud is increasing each day, especially in the recent two years following the COVID-19 pandemic. The sudden shutdowns of workplaces, institutions, and industries have increased the demand for cloud applications and services. The COVID-19 pandemic has highlighted the cloud opportunities to a large number of businesses that are expected to strengthen their utilization of the cloud even after the pandemic ends. A recent study by “Research and Markets”\footnote[4]{https://www.researchandmarkets.com/r/ys29pn} shows that the CC industry is expected to grow from \$371.4 Billion in 2020 to \$832.1 Billion by 2025. With this increase, a couple of important questions should be asked:  1) What is the extent of cloud scalability? and 2) Will the cloud be able to support the demand of users that reside in weakly connected areas? 

Cloud scalability is related to the number of customers that a specific cloud data center can serve. Several factors play an important role in scalability, such as the computing, storage, and bandwidth demands of the user applications that consume the services of the cloud data center. With the increase in the number of heavy applications that demand high processing, storage, bandwidth, or a combination of the three, the cloud is facing a scalability challenge. In the near future, it is expected that many organizations and institutions such as hospitals, universities, research centers, traffic control centers, etc., will be exploiting the cloud to run big data and AI applications that require specific quality of service (QoS) constraints. In addition, the drastic increase in the utilization of Internet of things (IoT) devices to collect and make use of various sensors’ readings has increased the demand for high cloud storage capabilities. Such applications will generate a huge amount of data each second and require frequent and high-complexity computing power to process this data continuously. With the increase in the scalability requirements, cloud providers will look for more and better suitable places to build cloud data centers that can support high processing complexity. The main issue here is that the devices of a cloud data center devour vast amounts of electricity that require a high-cooling process. The high-cooling demand of cloud data centers has driven cloud providers to apply costly solutions, such as deploying data centers deeply underwater or at the earth poles in order to make use of very low temperatures at these locations.

The second issue is related to the ability of the cloud to provide satisfiable QoS to user applications that are consumed at poor connectivity locations. Currently, there is still a large percentage of areas worldwide that suffer from poor Internet connectivity, especially in the developing and the least developed countries \cite{johnson_2021}. Many cloud services cannot be consumed from these locations because the QoS constraints of these services are not met in these locations (such as minimum bandwidth, maximum latency, etc.). Hence, it remains a challenge for cloud providers to provide their services with an acceptable QoS to users that reside in poorly connected areas. 

\subsubsection{Opportunities in Cloud-enabled HAPS}

The HAPS network can provide solutions and answers to the two raised questions with optimized design and implementation. First, the HAPS nodes can extensively enhance the cloud scalability by hosting new flying data centers. Exploiting the largely available physical space on the HAPS nodes to deploy powerful data centers will help cloud providers upscale their services and support more user applications. Since the atmospheric temperature at the HAPS elevation is naturally low, in the range of [-15\textsuperscript{o}C;-50\textsuperscript{o}C] \cite{kurt2020vision}, there is no need for too much energy for cooling. In addition, due to its location at a high altitude, the HAPS node can provide coverage to a wide ground area (around 500 km radius \cite{kurt2020vision}). With LoS links between the user devices and the HAPS node, the latter can offer cloud services with satisfiable QoS to users in areas that suffer from poor ground connectivity. Moreover, the HAPS network enables users to avoid several problems that typically occur when consuming cloud services via ground connections, such as the possibility of disconnection while offloading data due to poor channel conditions.

\begin{figure*}[!t]
\centering
\includegraphics[width=5.5in]{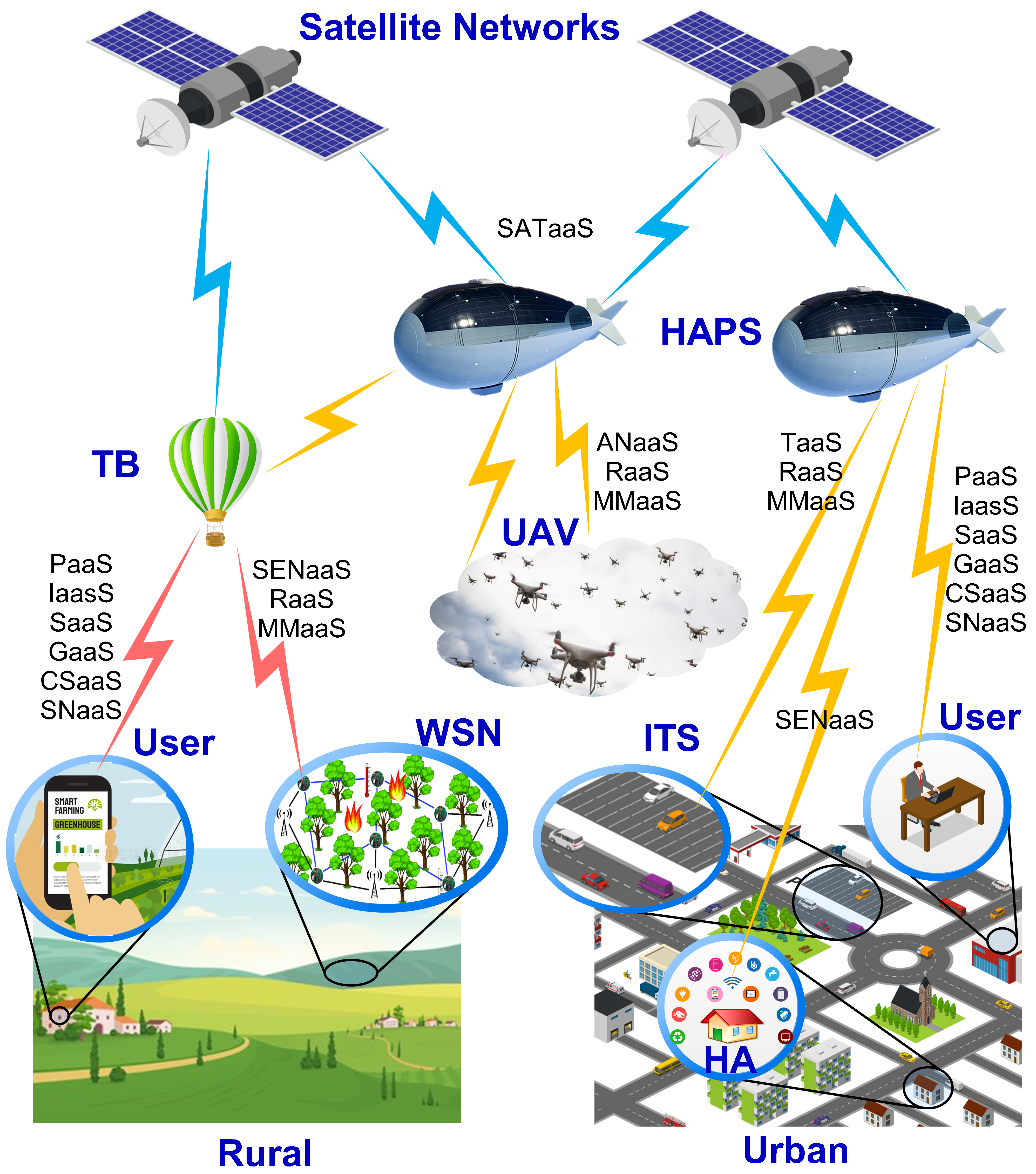}
\caption{Overview of the proposed services within the cloud-enabled HAPS.}
\label{FIG:allServices}
\end{figure*}

For these reasons, it is convenient that cloud providers deploy their services within the HAPS data centers. In addition to the novel HAPS-specific cloud services discussed in the following sections, we expect various traditional cloud services to be deployed on the HAPS nodes. Such services include platform as a service (PaaS), software as a service (SaaS), infrastructure as a service (IaaS), storage as a service (STaaS), and security as a service (SECaaS), among others. Offering such services to various users and businesses further satisfies urgent demands for higher scalability and rapid new cloud resources; it satisfies QoS requirements that ground cloud services fail to achieve. Figure \ref{FIG:allServices} provides an overview of the various HAPS cloud services that we discuss in this paper and the environments in which they are offered and consumed. These services are detailed in the next subsections.

\subsection{Satellite As a Service (SATaaS)}
\label{Sec:SATaaS}
\subsubsection{Overview}

Satellite services comprise a wide area of applications that can be categorized into three domains: Telecommunications, broadcasting, and data communications. Currently, there are more than 2,500 active satellites that orbit around the Earth, and the number is expected to significantly increase in the upcoming years. Several companies are investigating the use of large low earth orbit (LEO) satellite constellations to improve their services and reach more users; the HAPS network can be deployed as a middle layer between the ground and satellite networks. One of the main advantages of HAPS, as compared to LEO satellites, is their closer proximity to the ground level, and so connecting to them becomes much faster. Therefore, a HAPS constellation-based communication system can overcome the inherent high-latency problem of satellite networks \cite{kurt2020vision}. 

Recently, Microsoft announced Azure Orbital\footnote[5]{https://azure.microsoft.com/en-us/services/orbital/}, a service that offers to satellite operators the possibility to connect to and manage their satellite, process data, and control satellite activities directly from the cloud. Similarly, Amazon "AWS Ground Station" provides several satellite-related services, such as weather forecasting; surface imaging; communications; and video broadcasts\footnote[6]{https://aws.amazon.com/ground-station/}. Yao \textit{et al.} \cite{yao2020enabling} present a spatial cloud-based solution for analyzing and utilizing Big Earth Observational Data (BEOD), which are images that reflect specific characteristics in a time and space interval in the objective world. Kanev and Mirenkov \cite{kanev2011satellite} describe the "Satellite Cloud Computing" model, in which virtualized information resources from satellites are integrated within the cloud. The authors propose a general architecture of a cloud infrastructure with direct access to satellite data, and discuss the possibilities for bridging professional and home-based satellite receivers into an integrated satellite cloud computing framework. Finally, the authors in \cite{ujjwal2019cloud} describe how the computation, storage, and networking capabilities of cloud computing enable natural hazard modeling systems that utilize Satellite Remote Sensing to achieve accurate results in real-time.

\subsubsection{Opportunities in Cloud-enabled HAPS}
One of the main applications that the HAPS cloud framework can offer is satellite as a service (SATaaS) in which HAPS nodes act as the mediator between ground networks (including unmanned aerial networks or UAVs) and the satellite network. Current satellite services such as global navigation satellite system (GNSS), telephone, broadcasting, and monitoring services (weather, financial, environmental, etc.) can be offloaded from the satellites to the HAPS data centers and then offered to customers with better QoS, such as lower delay or better accuracy (for example, in localization services). In particular, the HAPS data centers can host and execute the server-side code of applications while continuously obtaining input data from various satellite networks. Furthermore, cloud customers can consume these services from the HAPS data centers using suitable CC partitioning and sharing models. In addition, the HAPS infrastructure and equipment could be used to improve the services whenever possible. For example, positioning sensors can be deployed on the HAPS nodes to improve the localization accuracy (much higher resolution than GNSS). Another important reason for offloading satellite services to the HAPS is enabling delay-sensitive applications to obtain the required information from the HAPS cloud service very fast, meeting the strict delay requirements. Such applications cannot afford the delay of obtaining information from satellites in traditional services. Therefore, deploying satellite services on the HAPS cloud, as opposed to consuming the service directly from the satellite network, allows application users to enjoy a higher QoS. 

\subsection{Sensor as a Service (SENaaS)}
\label{Sec:SENaaS}
\subsubsection{Overview}
A large number of IoT applications are currently offered as cloud services. IoT devices utilize various sensors that continuously read data from the environment and transmit it to IoT applications. Such IoT devices can be connected to a local server or to the cloud via gateways. IoT cloud platforms thus combine the powers of IoT devices and cloud computing in a single model that delivers an end-to-end service. Currently, there are more than 20 billion IoT devices connected to the Internet, and the number is expected to reach 38 billion in 2025\footnote[7]{https://www.statista.com/statistics/802690/worldwide-connected-devices-by-access-technology/}; there is an increasing need for gathering the big data obtained from such devices and processing it effectively via various applications. 

Cloud providers offer numerous IoT cloud platforms, hosting a wide range of IoT applications in automation, home security, agriculture, environmental monitoring, networking, etc. Examples of IoT cloud platforms include Oracle IoT, Google Cloud IoT, SAP IoT, Cisco Jasper, Altair SmartWorks, Ubidots, and SensorCloud. Also, there are special types of IoT cloud services utilized during specific events or situations such as emergencies and natural disasters. The latter usually use of advanced ML algorithms for predictive analysis and fast recovery (IBM DRaaS, Bluelock, etc.).

\subsubsection{Opportunities in Cloud-enabled HAPS}
The HAPS network is expected to offer new possibilities to current IoT cloud platforms and open the way for new types of platforms. A group of IoT sensors that monitor a particular place or environment could be connected to a HPAS gateway that frequently aggregates the sensors' readings and transmits them to the HAPS node. This setup is similar to the current setup in wireless sensor networks (WSNs), except that WSN gateways connect to the cloud via a base station or a dedicated wired connection \cite{mershad2020framework}. In some situations, it is hard to secure a reliable Internet connection for WSN gateways, especially when WSNs are deployed in rural locations, deep inside forests or deserts, or on top of high mountains. In such cases, dedicating a robust LoS connection between WSN gateways and HAPS nodes guarantees fast offloading of the WSN data to the HAPS cloud, and hence better QoS to users. In addition, HAPS nodes are expected to have high storage capacities \cite{kurt2020vision} that help various types of WSNs overcoming the limited storage problem in their local servers, thus enhancing these networks' applications performance. Furthermore, in scenarios where the WSN location is not covered by any HAPS node, tethered balloons (TBs) can be used to connect the WSN (and other ground users and networks) to the HAPS network. TBs were proposed in several works as a means for connecting multiple ground sources \cite{alsamhi2019performance,alzidaneen2019resource}. As shown in Fig. \ref{FIG:allServices}, in the absence of HAPS connectivity, TBs play an essential role in connecting the ground users and networks to the HAPS cloud. Several TBs can be placed in rural areas to connect to each other and connect via one or more TBs to the HAPS network. This solution is economically more feasible than placing HAPS nodes in rural areas.

\begin{figure}[!h]
  \begin{center}
  \includegraphics[width=3.5in]{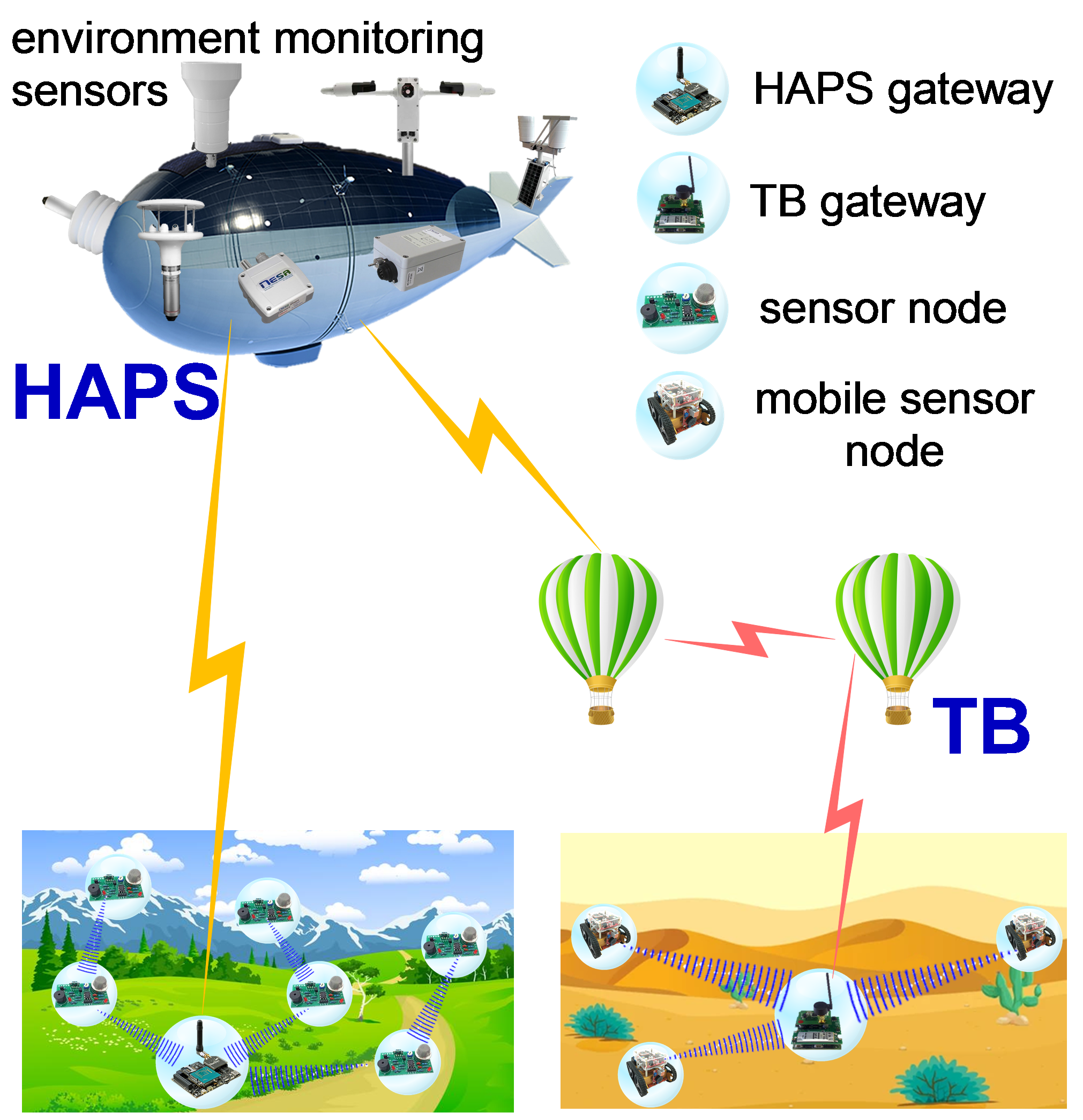}  
  \caption{Two SENaaS scenarios: 1) collecting data from ground sensors and sending them to the HAPS cloud via HAPS gateways or TBs, 2) monitoring the stratosphere via environment monitoring sensors installed within the HAPS node.}
  \label{fig:SENaaS}
  \end{center}
\end{figure}

HAPS can also open the way to new critical IoT cloud services. For example, HAPS nodes can be equipped with various types of sensors that monitor the environment and record the events that occur within the HAPS zone. The readings of such sensors can be directly transmitted to the HAPS node cloud server, processed by the ML algorithms, and then offered to the cloud users. The distinctive advantage of these types of HAPS cloud applications is that they allow the ground users to detect and/or predict important events deduced from the HAPS sensors' readings in real-time. Such features could help emergency and disaster teams build new models for natural disasters based on the sensor readings, which helps forecast future events more accurately. In addition, the readings of the HAPS sensors help scientists understand and discover new facts about the stratosphere. Fig. \ref{fig:SENaaS} illustrates the two SENaaS scenarios explained in this section. 

\subsection{Transportation as a Service (TaaS)}
\label{Sec:TaaS}
\subsubsection{Overview}
The research on vehicular cloud computing has flourished in the last few years, with a large number of papers proposing various models for integrating intelligent transport system (ITS) applications into the cloud \cite{sharma2020vanets, soyturk2016vehicular, mershad2013crown}. Such integration uses new features from the Internet of vehicles (IoV) to provide different cloud solutions for transportation companies. The authors in \cite{sharma2020vanets} provide a good overview of such ITS-based cloud services. In general, transportation cloud services can be offered to several types of cloud customers. For example, real-time vehicle tracking and vehicle health monitoring services are offered to transportation companies to monitor and manage their fleets of vehicles \cite{soyturk2016vehicular}. Furthermore, accident alert, road navigation, e-ticketing, and passenger entertainment services are offered to road passengers; smart parking services are offered to companies that require automatic parking solutions for their parking lots; road traffic management and traffic resolution services are offered to traffic control centers; disaster/emergency management services are offered to various emergency-related agencies and departments.

The actual implementation of vehicular cloud computing, however, is still limited. Very few cloud providers currently offer ITS-related services within their cloud platforms. For example, Amazon web services (AWS) provide several transportation solutions to related businesses. AWS services include traffic flow management, traffic congestion resolution, and disaster recovery\footnote[8]{https://aws.amazon.com/stateandlocal/transportation/}. Similarly, Google has recently started the connected vehicle platform as part of its cloud IoT core with services such as monitoring driving behaviors, predictive maintenance, and freight tracking\footnote[9]{https://cloud.google.com/solutions/designing-connected-vehicle-platform}. Netrepid\footnote[10]{https://www.netrepid.com/technology-strategies/transportation/} also offers customized transportation solutions based on the customer’s exact needs. These solutions include vehicle tracking and managing routes (for railroad and airline companies). Nevertheless, the number and range of vehicular cloud services are expected to increase in the next few years with the appearance of smart cities and the need to manage road data and operations in a fast and efficient manner. The latter opens the space for the cloud to provide many new applications to the ITS sector.

\subsubsection{Opportunities in Cloud-enabled HAPS}
\begin{figure}[!h]
  \begin{center}
  \includegraphics[width=3.5in]{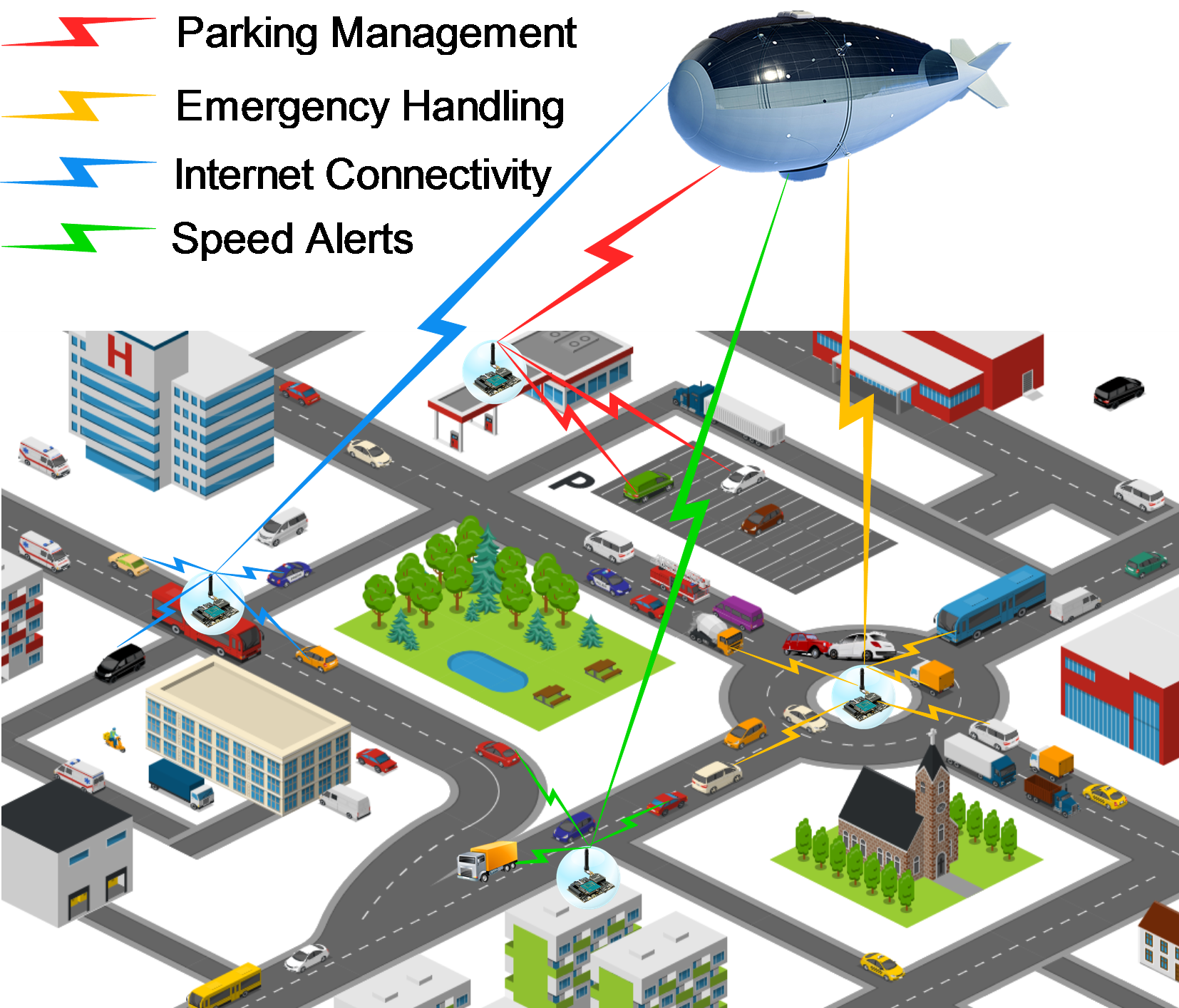}  
  \caption{Illustration of four ITS applications that utilize the HAPS cloud to offload data and offer services.}
  \label{fig:TaaS}
  \end{center}
\end{figure}

The most important advantage that the HAPS network can offer to the vehicular cloud is the ability to provide huge computational and storage capabilities. Future HAPS architectures are expected to support data acquisition, computing, caching, and processing in diverse application domains \cite{kurt2020vision}. Several recent works emphasized the fact that vehicular networks in future smart cities are expected to generate continuous streams of big data that require continuously increasing storage and very fast processing \cite{cheng2018big}, \cite{soleymani2020authentication}. Various ITS applications could be partially or entirely offloaded to the HAPS network and offered as cloud services to ITS administrators and customers. Consequently, various vehicular network nodes could offload their data to the HAPS cloud via multiple strategies. For example, some roadside units (RSUs) could be integrated with HAPS communication modules to establish a high-speed LoS connection to the HAPS network and support a high data rate for vehicular big data offloading. These RSUs can be deployed in high locations such as the top of buildings to enhance their connectivity with the HAPS network. Using such setups, the vehicular network can make use of the ground RSU-to-RSU (also denoted as infrastructure to infrastructure or I2I) connections to aggregate and transfer various ITS applications' data to RSUs, which in turn transmit the data to the HAPS network. Another possible solution is to equip certain types of vehicles that act as mobile gateways (such as buses) with HAPS communication modules. Such strategies create fast connection lines between the vehicular and the HAPS networks, enabling the former to use the latter's storage and processing capabilities. Fig. \ref{fig:TaaS} illustrates four ITS applications that can be offered to ITS customers via the HAPS cloud. 

\subsection{Aerial Network as a Service (ANaaS)}
\label{Sec:ANaaS}

\subsubsection{Overview}
Like vehicular cloud computing, cloud computing services for UAV networks are expected to become a hot research field in the near future. Some recent papers proposed various frameworks for integrating UAVs with cloud platforms. For example, reference \cite{luo2015uav} discusses a model for offloading the UAVs sensors’ data to cloud servers. Their framework mainly focuses on disaster and emergency scenarios. In particular, a UAV node captures videos of its environment and passes them to a context scheduler that analyzes the video frames based on several context data such as location, spatial and temporal data, and control commands in order to decide on the parts of the video data that should be offloaded to a cloud server. The authors in \cite{pinto2019framework} further propose a combined cloud-fog architecture for UAVs in which several UAV nodes are chosen as fog coordinators. Such coordinators collect data from other UAV nodes, analyze and partition the data based on the application requirements, process the data locally or cooperate with cloud servers to produce the application results, and coordinate the operations of other UAV nodes to satisfy the application demands.

The work in \cite{mahmoud2015toward} outlines several cloud services that UAV networks can offer. These services include UAV IaaS, which utilizes the UAV hardware components such as payloads, sensors, actuators, internal memory, and processor. They also include UAV PaaS, in which the client can make the UAV nodes read and send specific data using their sensors or perform an action using certain actuators. Furthermore, the UAV SaaS enables cloud users to request certain UAV missions, such as commanding UAVs to spray crops for a specific agricultural area. The authors in \cite{loke2015internet} outline various challenges that face UAV cloud applications and scenarios. Such services involve managing the big data generated by UAV nodes, optimizing the UAV operations for multiple cloud services, maintaining the situation awareness of UAV nodes, ensuring the reliability of UAV collected data, enabling scheduling to partition the cloud service among UAV nodes, and avoiding various security attacks.

\subsubsection{Opportunities in Cloud-enabled HAPS}
\begin{figure}[!h]
  \begin{center}
  \includegraphics[width=3.5in]{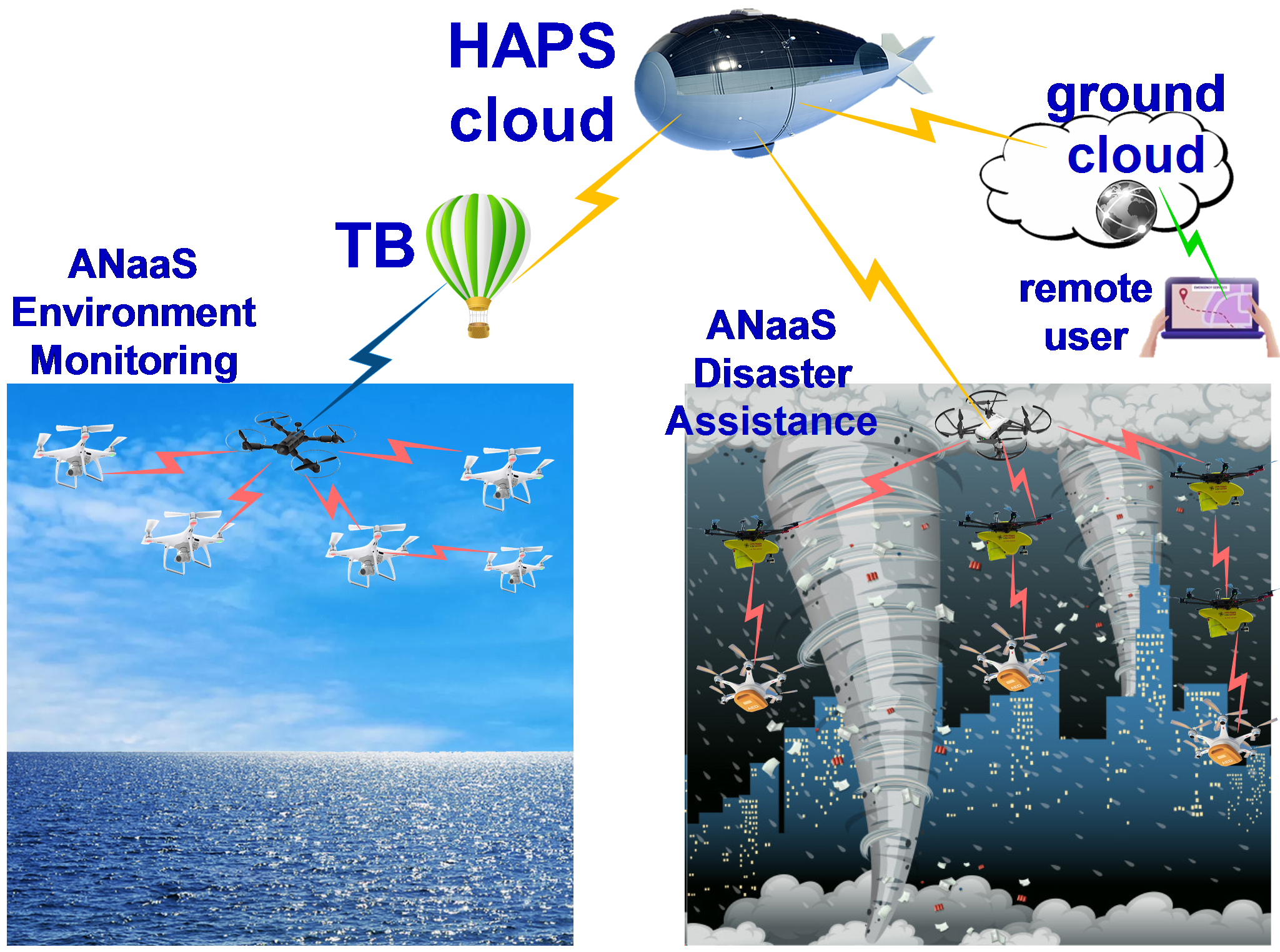}  
  \caption{Two scenarios in which the UAV network can make use of the HAPS cloud: 1) rural locations where ground connectivity is not available, and 2) emergency situations where ground connectivity is weak or off due to unusual conditions such as natural disasters.}
  \label{fig:ANaaS}
  \end{center}
\end{figure}

UAV nodes (drones and aerial BSs) usually connect to the Internet via ground gateways in urban areas. However, in rural areas, seas, oceans, or deserts, the drones might not have the means to connect via a ground gateway; the HAPS nodes can act as the gateway. Similar to the discussion in the previous section, we note that UAV nodes can be equipped with HAPS communication modules that enable them to establish a LoS connection with a HAPS node. Consequently, the UAV nodes with a HAPS connection can collect, filter, and aggregate various data from other UAV nodes and offload this data to the HAPS cloud data center. Cloud users then consume the UAV services by connecting to the HAPS data center and executing the corresponding service. This setup enables cloud users to connect to UAVs positioned in weakly connected or disconnected locations. UAV cloud services enable the cloud users to analyze the UAV sensors’ readings to monitor or study the UAV area and command the UAV nodes to move to certain locations or perform specific tasks. In addition, UAVs deployed in weakly connected or disconnected areas enable users to connect to the cloud via the UAV-HAPS connection. Even in scenarios where the UAV has ground connectivity, UAV applications that require large storage and processing overheads can still offload their data and code to the HAPS cloud instead of the ground cloud due to the advantages of the former, such as better connection and channel conditions, lower latency, etc. These advantages mostly appear when the UAVs encounter obstacles (emergencies that affect the ground infrastructure such as fire or floods, for example) or are deployed in obstructed areas. Fig. \ref{fig:ANaaS} illustrates two examples in which the HAPS cloud can assist the UAV nodes in offering cloud services.

\subsection{Routing as a Service (RaaS)}
\label{Sec:RaaS}
\subsubsection{Overview}
Routing in wireless networks is the process of calculating the path that a packet should follow when sent from a source to a destination node. Different types of wireless networks require different routing protocols and strategies that best suit the characteristics of a network. For example, WSNs and IoT networks utilize multi-hop routing paths to forward a packet from one sensor node to another until the packet reaches one of the network gateways. Vehicular networks may depend on fixed infrastructures such as roadside units connected to the Internet to help in the routing process. Towards this end, two main types of communications, namely, vehicle-to-vehicle (V2V) and vehicle-to-infrastructure (V2I), could be used in vehicular routing protocols. Similarly, UAVs rely on UAV nodes equipped with ground communication modules to connect to ground gateways. Since only some UAV nodes could have ground connectivity, the remaining nodes need to find multi-hop paths, connecting to one or more ground-connected UAV nodes to route packets to ground networks.

Although different routing protocols exist for different wireless networks, the next generation of routing protocols will arguably utilize network virtualization techniques such as software-defined networking (SDN) and network functions virtualization (NFV) for better routing efficiency. Several SDN-based and NFV-based routing protocols have already been proposed for various wireless networks. For example, the authors in \cite{younus2019proposition} and \cite{ouhab2020energy} propose SDN-based routing protocols for WSNs and large-scale IoT networks, respectively. The SDN-based routing protocols for WSNs outperform other protocols in terms of energy efficiency and reducing the power consumption of WSN nodes \cite{younus2019proposition}. Furthermore, the SDN routing protocol in \cite{ouhab2020energy} outperforms other routing protocols for IoT networks in terms of supporting higher network scalability and larger-size networks. The works in \cite{mershad2020surfer, mershad2020utilizing, mershad2020block} prove that utilizing the SDN combined with RSUs is more efficient than traditional routing methods for vehicular networks. Finally, the authors in \cite{secinti2018sdns} and \cite{xiong2019sdn} propose two similar SDN-based routing protocols for UAVs in which multiple paths are created between two UAV nodes. These protocols focus on selecting the routing strategy according to the required QoS parameter, such as selecting the shortest path to decrease the end-to-end delay or sending the packet on multiple paths to increase the packet delivery ratio. 

\subsubsection{Opportunities in Cloud-enabled HAPS}
Future wireless network routing protocols can utilize HAPS-deployed SDN controllers to process routing data from network nodes, generate routing decisions (such as OpenFlow rules), and send routing updates to wireless network nodes (to update the routing tables). Such protocols have two main advantages compared to traditional ground-based SDNs. The first advantage is that the HAPS nodes can cover a wide ground area (footprint). Consequently, wireless networks that do not have a fast connection to a ground cloud server with high processing capabilities can switch to using the HAPS cloud for its high processing capacity and fast connection. This is especially true for very large wireless networks that contain tens or hundreds of thousands of nodes, such as huge WSNs or IoVs that generate continuous streams of routing data that should be sent to the SDN controller. Note that there should be an efficient filtering and aggregation procedure in such networks to enable the wireless network nodes to filter redundant data and group important data before sending them to the HAPS SDN controller. 

The second advantage of such protocols is related to the fact that SDN switches require stable and reliable connections to the SDN controller, which is something that cannot always be guaranteed in current SDN architectures for wireless networks. Many reasons cause connection loss in these networks, such as high interference and/or noise, the existence of blocking objects (WSNs that are deployed in rural and environment-changing areas), and high mobility of highway vehicles which does not leave vehicles enough time to send packets to the RSU. Suppose these nodes are equipped with a HAPS communication module or connected to other nodes with HAPS communication modules (such as TBs). In that case, we can avoid the network isolation problem that occurs when one or more SDN switches cannot reach the SDN controller.

\subsection{Other Cloud Services}
\label{Sec:otherServices}
In addition to the cloud services mentioned in the previous sections, other services are expected to be implemented and offered within the HAPS cloud as the HAPS network matures and more HAPS constellations are built. We provide some examples of cloud services that could make use of a powerful HAPS architecture to provide a high QoS in this section.

\subsubsection{Mobility Management as a Service (MMaaS)}
\label{Sec:MMaaS}
A vital cloud service related to mobile wireless networks that the HAPS cloud system can offer is mobility management as a Service (MMaaS). Mobility is an important characteristic that provides great advantages to networks, but at the same time creates several problems, such as dynamic topology and intermittent connectivity, complex modulation and coding, handoff, road traffic congestion (in vehicular networks), etc. Several research works proposed mobility management systems for mobile networks. For example, the authors of \cite{vijayarangam2018vehicular} propose a mobility model that is implemented within a vehicular cloud platform. The proposed model makes use of free flow and queuing-up movements of vehicles to suggest an alternate route for the driver which is free from traffic. Another model for calculating and managing the trajectories of drones in UAVs is proposed in \cite{nguyen2021autonomous}. Here, the drone routes are estimated via the global GPS reference system with supporting GIS mapping, coupled with autonomous conflict detection, resolution, safe drone following, and formation flight options.

With HAPS-enabled MMaaS, a mobility management cloud server can execute ML algorithms that analyze various data to be sent from the mobile nodes to the HAPS cloud, such as geographic location, movement direction, speed, and acceleration. The server can then build a movement profile for each mobile node by analyzing the mentioned parameters during a certain time period; such data can be used to predict future movements and network changes. The server can dynamically re-execute the ML algorithms when the inputs change (upon nodes entering and leaving, speed and acceleration changes, weather changes, etc.). MMaaS can overcome many of the problems mentioned above by producing a future movement trajectory for each mobile node to guarantee that the network continues to operate normally. For example, if the nodes’ movement profiles indicate that part of the WSN will be isolated from the network if the nodes retain their movement pattern, the MMaaS can adjust the future movement paths of one or more nodes to guarantee that no sub-network isolation occurs. Another example is vehicular traffic control. Although many efforts have been made in road traffic control and traffic congestion avoidance, road traffic is still causing huge problems, especially in high-urban areas. HAPS nodes can combine the data that we mention in this paragraph with high-resolution images that they can take and that show the exact concentration of vehicles on the roads, the exact locations of accidents, and the existence of temporary obstacles on certain roads. When combining and analyzing all this data, the HAPS nodes can predict the future traffic patterns of the vehicles and advise each driver to follow a specific path to resolve expected future traffic congestion and balance the traffic on all the paths to a particular destination. 

\subsubsection{Gaming as a Service (GaaS)}
\label{Sec:GaaS}
Cloud gaming has been drawing increasing attention in the cloud and game industries. Several research works discussed the benefits of cloud-gaming platforms in improving the performance of end-user games and creating a network architecture for a high QoS multiplayer cooperative cloud gaming \cite{chuah2014cloud, cai2018toward, gao2019cost}. With HAPS in the picture, cloud gamers' devices can be equipped with high-speed HAPS communication modules that allow the gamers to connect to the HAPS cloud from their devices. Using this setup, players can run the front-end of the game on their devices, and the devices could connect to the back-end of the game application on the nearest HAPS node. In addition, the HAPS nodes can collaborate to run the games in parallel between large groups of gamers and balance the load within the HAPS network. This setup enables gamers to play various single and multiplayer games that are hosted on the HAPS cloud. Several cloud platforms, such as Vortex\footnote[11]{https://vortex.gg/}; Shadow\footnote[12]{https://shadow.tech/}; and Amazon Luna\footnote[13]{https://www.amazon.com/luna/}, are currently offering similar services on the ground cloud. The main advantage that HAPS offer to cloud gamers is that each HAPS node has a large footprint, allowing mobile gamers to guarantee a stable connection to the same HAPS node for a large period of time and avoid the frequent handoff problem that decreases the QoS of the current cellular systems.

\begin{table*}[htp]
\caption{Summary of the proposed HAPS-cloud services.}
\label{Tab:ServicesSummary}
\centering
\begin{tabular}{p{0.06\linewidth}p{0.28\linewidth}p{0.28\linewidth}p{0.28\linewidth}}
\hline
\textbf{HAPS Cloud Service}	& \textbf{Existing Services} &	\textbf{Advantages} & \textbf{Limitations} \\
\hline
SATaaS	& 
\begin{itemize}[leftmargin=*,noitemsep,partopsep=0pt,topsep=0pt,parsep=0pt]
    \item Azure Orbital
    \item AWS Ground Station
	\item Spatial Cloud for Big Earth Observation Data \cite{yao2020enabling}
	\item Satellite Cloud Computing \cite{kanev2011satellite}
	\item Cloud-based natural hazard modeling systems \cite{ujjwal2019cloud}
\end{itemize}
&
\begin{itemize}[leftmargin=*,noitemsep,partopsep=0pt,topsep=0pt,parsep=0pt]
	\item Lower delay
    \item More accuracy
    \item Faster detection and response (natural hazards)
    \item Better quality (media and TV)
\end{itemize}
&
\begin{itemize}[leftmargin=*,noitemsep,partopsep=0pt,topsep=0pt,parsep=0pt]
    \item Less coverage per node
    \item Frequent maintenance and update requirements
    \item Possibility of disconnection to the HAPS network
\end{itemize}
 \\
\hline
SENaaS
&
\begin{itemize}[leftmargin=*,noitemsep,partopsep=0pt,topsep=0pt,parsep=0pt]
    \item Oracle IoT
    \item Google Cloud IoT 
	\item SAP IoT 
	\item Cisco Jasper
	\item Altair SmartWorks
	\item Ubidots
	\item SensorCloud	
\end{itemize}
&
\begin{itemize}[leftmargin=*,noitemsep,partopsep=0pt,topsep=0pt,parsep=0pt]
    \item Ability to deploy in rural and isolated areas
	\item Monitoring the environment at the stratosphere layer
	\item Meeting the time requirements of delay-sensitive applications
\end{itemize}	
&
\begin{itemize}[leftmargin=*,noitemsep,partopsep=0pt,topsep=0pt,parsep=0pt]
    \item Requiring dedicated HAPS gateways
	\item Requiring TBs in areas not covered by the HAPS network
\end{itemize}
\\
\hline
TaaS
&
\begin{itemize}[leftmargin=*,noitemsep,partopsep=0pt,topsep=0pt,parsep=0pt] 
    \item AWS Automotive
    \item Google Connected Vehicle Platform
    \item Netrepid Transportation Cloud
    \item AzureITS
\end{itemize}
&

\begin{itemize}[leftmargin=*,noitemsep,partopsep=0pt,topsep=0pt,parsep=0pt]
    \item Computational and storage capabilities for ITS big data applications
    \item LoS connection
    \item Ability to consume services from mobile devices
    \item Few HAPS nodes can cover the whole vehicular network	
\end{itemize}
&
\begin{itemize}[leftmargin=*,noitemsep,partopsep=0pt,topsep=0pt,parsep=0pt]
    \item Requiring dedicated HAPS gateways
    \item Disconnections in underground and hidden areas
\end{itemize}
\\
\hline
ANaaS
&
\begin{itemize}[leftmargin=*,noitemsep,partopsep=0pt,topsep=0pt,parsep=0pt] 
    \item Disaster Sensing \cite{luo2015uav}
    \item Object recognition \cite{pinto2019framework}
    \item UAV IaaS, UAV PaaS, UAV SaaS \cite{mahmoud2015toward}
    \item Drone as a Service \cite{loke2015internet}
\end{itemize}
&
\begin{itemize}[leftmargin=*,noitemsep,partopsep=0pt,topsep=0pt,parsep=0pt] 
    \item Ability to deploy in rural and isolated areas
    \item LoS connection
    \item Connecting users in ground-disconnected areas or conditions
    \item Computational and storage capabilities for UAV big data applications	
\end{itemize}
&
\begin{itemize}[leftmargin=*,noitemsep,partopsep=0pt,topsep=0pt,parsep=0pt] 
    \item Requiring dedicated HAPS gateways
    \item Requiring TBs in areas not covered by the HAPS network
    \item Higher End-to-End delay between the user and the UAV (in some cases)
\end{itemize}
\\
\hline
RaaS
&
\begin{itemize}[leftmargin=*,noitemsep,partopsep=0pt,topsep=0pt,parsep=0pt]
    \item Energy-Aware Multihop Routing Protocol (EASDN) \cite{younus2019proposition}
    \item Routing Protocol for Low-Power Lossy Networks (RPL) \cite{ouhab2020energy}
    \item Secure Routing Protocol for Internet of Vehicles (SURFER) \cite{mershad2020surfer}
    \item SDN-based UAV (SD-UAV) \cite{secinti2018sdns}
\end{itemize}
&
\begin{itemize}[leftmargin=*,noitemsep,partopsep=0pt,topsep=0pt,parsep=0pt]
    \item Ability to group all routing data from a huge network at a single HAPS node
    \item Stable and reliable connection between routing switches and controller
\end{itemize}
&
\begin{itemize}[leftmargin=*,noitemsep,partopsep=0pt,topsep=0pt,parsep=0pt]
    \item Requiring dedicated HAPS gateways
    \item Requiring TBs in areas not covered by the HAPS network
    \item Disconnections in underground and hidden areas
    \item Limited physical space within the HAPS node to support routing hardware
\end{itemize}
\\
\hline
MMaaS
&
\begin{itemize}[leftmargin=*,noitemsep,partopsep=0pt,topsep=0pt,parsep=0pt]
    \item Free-flow model, Queuing- up model \cite{vijayarangam2018vehicular}
    \item Trajectory-following model, Drone-following model \cite{nguyen2021autonomous}
\end{itemize}
&
\begin{itemize}[leftmargin=*,noitemsep,partopsep=0pt,topsep=0pt,parsep=0pt]
    \item Computational and storage capabilities for traffic and mobility big data applications
    \item Meeting the strict delay requirements of mobility management applications
\end{itemize}
&
\begin{itemize}[leftmargin=*,noitemsep,partopsep=0pt,topsep=0pt,parsep=0pt]
    \item Requiring dedicated HAPS gateways
    \item Requiring TBs in areas not covered by the HAPS network
    \item Real-time connection is not always guaranteed
\end{itemize}
\\
\hline
GaaS
&
\begin{itemize}[leftmargin=*,noitemsep,partopsep=0pt,topsep=0pt,parsep=0pt]
    \item Vortex
    \item Shadow
    \item Amazon Luna
\end{itemize}
&
\begin{itemize}[leftmargin=*,noitemsep,partopsep=0pt,topsep=0pt,parsep=0pt]
    \item Meeting the strict delay requirements of gaming applications
    \item Stable LoS connection
    \item No Handoff
\end{itemize}
&
\begin{itemize}[leftmargin=*,noitemsep,partopsep=0pt,topsep=0pt,parsep=0pt]
    \item Requiring dedicated connection between the user device and the HAPS node
    \item Disconnections in underground and hidden areas
\end{itemize}
\\
\hline
SNaaS
&
\begin{itemize}[leftmargin=*,noitemsep,partopsep=0pt,topsep=0pt,parsep=0pt]
    \item Social Cloud \cite{chard2011social}
    \item Healthcare cloud services \cite{sandhu2016smart}
    \item Social network educational services \cite{thaiposri2015enhancing}
    \item Multimedia as a Service \cite{nan2014distributed}
\end{itemize}
&
\begin{itemize}[leftmargin=*,noitemsep,partopsep=0pt,topsep=0pt,parsep=0pt]
    \item Lower delay
    \item Computational and storage capabilities for SN applications
    \item Connection to ground SN via the HAPS
\end{itemize}
&
\begin{itemize}[leftmargin=*,noitemsep,partopsep=0pt,topsep=0pt,parsep=0pt]
    \item Requiring dedicated connection between the user device and the HAPS node
    \item Disconnections in underground and hidden areas
\end{itemize}
\\
\hline
CSaaS
&
\begin{itemize}[leftmargin=*,noitemsep,partopsep=0pt,topsep=0pt,parsep=0pt]
    \item Crowdsourcing Reference Model \cite{murturi2015reference}
    \item Crowdsourcing Social Enterprise Service \cite{tung2017crowdsourcing}
\end{itemize}
&
\begin{itemize}[leftmargin=*,noitemsep,partopsep=0pt,topsep=0pt,parsep=0pt]
    \item Computational and storage capabilities for crowdsourcing applications
    \item Ability to reach users in disconnected areas 
\end{itemize}
&
\begin{itemize}[leftmargin=*,noitemsep,partopsep=0pt,topsep=0pt,parsep=0pt]
    \item Higher delay than traditional crowdsourcing applications (in some cases)
\end{itemize}
\\
\hline
\end{tabular}
\end{table*}

\subsubsection{Social Network as a Service (SNaaS)}
\label{Sec:SNaaS}
Social networks (SN) have seen enormous expansion, with millions of Internet users actively engaging across various social networking platforms. Many businesses have started utilizing social networks as a means to promote their services, reach new customers, and enhance their connections with current customers. Instead of linking people based on social relations, businesses create interactive societies that group persons based on common business desires or practices. The social cloud is proposed in \cite{chard2011social}, enabling cloud users to share heterogeneous resources from multiple cloud services within the context of a social network. The social cloud comprises a social marketplace that utilizes both social and economic protocols to manage the sharing of resources and facilitate trading between users. In general, social networking cloud services have been used in the context of many fields, such as healthcare (\cite{sandhu2016smart}), education (\cite{thaiposri2015enhancing}), sharing of cloud resources (\cite{caton2014social}), and media (\cite{nan2014distributed}). In the future, HAPS nodes can host cloud data centers dedicated to social network cloud applications. The HAPS social network data center could include servers for implementing and managing the SN security, storage, applications, communications, backup, policies, etc. Similar to what we stated in Sec. \ref{Sec:GaaS}, the devices of the SN users would use a dedicated HAPS communication module to connect to the HAPS network, log in to the SN, and consume the desired services. Each HAPS node will connect to other HAPS nodes and to the ground data centers of the SN to enable its users to utilize various resources within the SN. The HAPS SN can offer several advantages as compared to current cloud social services, such as extra computational power, lower user-to-cloud and user-to-user delays (most probably in rural and weak connectivity areas, where users can connect to the HAPS either directly or via TBs), and better channel conditions.

\subsubsection{Crowdsourcing as a Service (CSaaS)}
\label{Sec:CSaaS}
Crowdsourcing is the practice of involving a group of people for a common target, such as voting, sharing information or opinions, or performing specific tasks via websites, social networks, and smartphone apps. It enables businesses to farm out work to individuals anywhere around the globe, and hence get access to a wide collection of skills and abilities without suffering the expenses of in-house employees. The new concept of crowdsourcing as a service enables companies to leverage cloud users and resources to execute crowdsourcing activities. The work in \cite{murturi2015reference} describes a reference model for cloud-based crowdsourcing. Furthermore, the work in \cite{tung2017crowdsourcing} proposes a system that allows social network users to post ideas, share resources, and engage in crowdsourcing projects, enabling businesses to seek crowdsourcing opportunities, collect, connect, and analyze users' data and resources. A similar framework in \cite{fuller2021crowdsourcing} regulates various operations within the crowdsourcing service, such as framing the crowd challenge, dealing with IP rights, managing the crowd, defining and handling incentives, and correctly integrating the vast input into innovation projects.

When the HAPS network matures, and people start to have HAPS communication modules within their devices, services such as CSaaS will be offered from the HAPS cloud and will become desirable for many companies. For example, when deployed in the HAPS cloud, the CSaaS enables businesses to create and host their crowdsourcing projects and lists within the HAPS data center, thus assisting individuals in participating in crowdsourcing tasks or subscribing to crowdsourcing lists. CSaaS could be implemented as both a push and a pull service. In some projects, the organization can select one or more crowdsourcing lists to send crowdsourcing tasks, such as voting, answering questions, or filling forms, with corresponding incentives (push service). In other cases, any subscriber can select a crowdsourcing task or project to take part in (pull service). The CSaaS subscriber can receive the crowdsourcing tasks via email or a social media message, and he/she can then reply to the crowdsourcing server with his/her credentials to receive the incentive. Organizations collect and analyze answers by utilizing data analytics tools within the CSaaS. HAPS nodes provide the middleware between crowdsourcing organizations and users by hosting lists, tasks, and answers, providing connectivity and message transfer, and executing applications.

\begin{table}[t]
\centering
\caption{Requirements for the various HAPS cloud services.}
\label{Tab:requirements}
\begin{tabular}{p{0.5\linewidth} | p{0.4\linewidth}}
\hline
\textbf{Requirement} & \textbf{Services}
\\ \hline
High-speed connection between the HAPS node and the service source/consumer & IaaS, SaaS, PaaS, SATaaS, SENaaS, TaaS, ANaaS, RaaS, MMaaS, GaaS, SNaaS
\\ \hline
Reliable connection between the HAPS node and the service source/consumer & IaaS, SaaS, PaaS, SATaaS, SENaaS, TaaS, ANaaS, RaaS, MMaaS, GaaS, SNaaS, CSaaS \\ 
\hline
High data-rate connection between the HAPS node and the service source/consumer & IaaS, PaaS, SATaaS, SENaaS, TaaS, ANaaS, RaaS, MMaaS, GaaS, SNaaS \\
\hline
Secure connection between the HAPS and the service source/consumer and secure processing/storage within the HAPS node & IaaS, SaaS, PaaS, SATaaS, SENaaS, TaaS, ANaaS, RaaS, MMaaS, GaaS, SNaaS, CSaaS \\
\hline
Dedicated HAPS gateway that connects the service’s source network to the HAPS & SENaaS, TaaS, ANaaS, RaaS, MMaaS \\
\hline
Dedicated communication module for connecting the HAPS to specific networks such as satellites or TBs & SATaaS, SENaaS, ANaaS \\
\hline
Dedicated connection between the consumer’s device and the HAPS & GaaS, SNaaS \\
\hline
Sufficient physical space within the HAPS node & IaaS, PaaS, GaaS, SNaaS \\
\hline
High processing capability by the HAPS & SaaS, PaaS, SENaaS, TaaS, ANaaS, RaaS, MMaaS, GaaS \\
\hline
High storage capacity within the HAPS & SENaaS, TaaS, ANaaS, RaaS, MMaaS, GaaS, SNaaS, CSaaS \\
\hline
Intra-HAPS collaboration between the HAPS nodes to execute/offer the service & SATaaS, ANaaS, RaaS, MMaaS, GaaS, SNaaS, CSaaS \\
\hline
Tethered Balloons (TBs) for connecting the service source/consumer with the HAPS & SENaaS, ANaaS, RaaS, MMaaS \\
\hline
Specialized hardware installed within the HAPS nodes for service-specific requirements (such as sensors, SDN controller, etc.) & SATaaS, SENaaS, RaaS \\
\hline
Support for real-time data transmission and sharing & SATaaS, SENaaS, TaaS, ANaaS, RaaS, MMaaS, GaaS \\
\hline
\end{tabular} 
\end{table}

A comparison between the services that we propose in this section for the HAPS-cloud platform and similar services that are offered in traditional or mobile cloud platforms is depicted in Table \ref{Tab:ServicesSummary}. For each service described in this section, we state in Table \ref{Tab:ServicesSummary} the existing similar services, the advantages that the proposed service will have and its limitations as compared to the existing services. In general, the HAPS-cloud platform will have unique features that will distinguish it from existing systems. The line-of-sight connection that will exist between the ”data source”/“consumer’s device” and the HAPS node makes the speed of offloading/consuming the application very high, which is significantly important for many cloud applications. In addition, the ability to deploy HAPS nodes over isolated and extremely remote areas enables the cloud providers to reach new cloud customers and opens the way for new types of cloud services. Moreover, the large footprint of a HAPS node makes it cover a wide ground area, which enables it to manage networks that span an enormous geographical zone and allows the nodes in these networks to make use of the computational and storage capabilities within the HAPS node to execute big data computationally-extensive applications (such as routing, ITS, social networks, etc.). Finally, the strategic location of the HAPS and its ability to communicate with several types of networks makes it ideal for connecting cloud users in these networks together.

We conclude this section by describing in Table \ref{Tab:requirements} the various requirements that are essential for the successful deployment of the proposed cloud services within the HAPS. For each requirement, we state the services that will crucially need the requirement. Some requirements, such as the first four in the table, are very important for most cloud services. Other requirements are needed by specific services. For example, only GaaS and SNaaS require a dedicated connection between the user's device and the HAPS node. This is due to the fact that these services are mostly consumed from mobile devices and require a stable high data-rate connection. In addition, some services, such as SENaaS and ANaaS, may require a middle layer between the data source and the HAPS, since the data collected by these services could be obtained from devices in rural or isolated areas that are not covered by the HAPS network. Hence, TBs will be used to obtain the data from these devices and send it to the HAPS, as described earlier. Another set of services, such as SATaaS, SENaaS, RaaS, require dedicated hardware to be installed within the HAPS nodes. For example, SENaaS needs environment monitoring sensors, SATaaS requires a navigation system, and RaaS entails dedicated routers (such as SDN controllers) to route packets between ground nodes.

\section{Challenges}
\label{Sec:challenges}
The real assessment of the cloud-enabled HAPS remains a strong function of the physical challenges of their practical deployment. Such challenges, in essence, stem from the HAPS nature as flying networks, their high power requirements, and the conditions of the stratosphere and its surrounding layer, all of which have to be factored in while designing such networks. This section entails some of the considerations which would arise while integrating cloud computing capabilities into HAPS, and discusses their potential solutions, whenever possible.

\subsection{Energy Requirements}
The high energy requirement needed to empower HAPS is considered a fundamental challenge in HAPS research in general \cite{kurt2020vision}. This is mainly because HAPS missions as aerostatic flying networks require long endurance, which is in the order of months or years. Such prolonged operation limits the use of classical energy sources, such as traditional aviation fuel \cite{Grace2011} or ground beams supplies \cite{Schlesak1988}, especially given the lengthy constraints on HAPS propulsion and electrical supply requirements, as well as the surrounding environmental considerations. In conventional HAPS deployment, solar power coupled with energy storage is considered essential to empower HAPS. This is primarily due to the absence of clouds in the stratosphere, which enables utilizing solar energy effectively. Arum \textit{et al.} \cite{arum2020energy} investigate the feasibility of utilizing solar power for long endurance, semi-permanent HAPS missions. The results show that the services of a HAPS node can be provided for a duration of 15–24 hours/day with wingspans ranging between 25–35 m. Other recent energy supply initiatives propose utilizing hydrogen power as a means to provide a long HAPS endurance and better environmental impact \cite{cambridgeconsultants}.

In the particular context of cloud-enabled HAPS proposed in our paper, the energy considerations become even more pronounced, especially given the computational capability requirements at the centralized processors available at the HAPS level. More precisely, the HAPS system envisioned in this paper amalgamates both powerful processing capabilities and strong connectivity premises, all while ensuring a prolonged, reliable, and effective system operation. This paradigm shift in HAPS operation turn HAPS into a super-macro base station (SMBS), with a multitude of stringent energy requirements \cite{Alam2021}. Such challenging aspects of future HAPS deployment would require sophisticated energy optimization and resource allocation schemes, so as to determine the proper energy resources under both short-term and long-term considerations (e.g., environmental impacts, deployment costs, weather forecasts, etc.), as further discussed in the future research directions of the paper.

\subsection{Interactions between HAPS and Surrounding Layers}
The cloud-enabled HAPS key infrastructure proposed in this paper is expected to corroborate the multi-layered integrated space-air-ground networks (also known as vertical heterogeneous networks (VHetNets)). Such networks are a mere abstraction of integrated satellites at the GEO-layer (i.e., GEO-satellites), LEO-layer (i.e., cubesats), stratospheric layer (i.e., HAPS and hot-air balloons), and aeronautical layer (i.e., low-altitude platform systems (LAPS) and UAVs), all inter- and intra-connected for better serving the ground-level communications infrastructures. Not only would the end-to-end utility function of such VHetNets depend on their highly coupled interactions and mutual dynamics, but it is also a strong function of the ever-changing environmental conditions across the different layers. For example, for systems where the downlink transmission from the upper satellites layers to the HAPS are done via FSO links, the system performance becomes a strong function of the impacts of component failure and atmospheric conditions, e.g., turbulence, fog, and rain \cite{Sharma2016, Zeidini2020}. Likewise, the connection from the HAPS layer to the ground level base stations becomes a function of the nature of the backhaul (RF or FSO) and its surrounding environmental conditions. For example, Grace \textit{et al.} \cite{grace2001providing} study the effect of the HAPS node displacement on the received signal power. They show that the antenna type (fixed or steerable) at the ground station greatly affects the system performance when the HAPS node is displaced from its location. The authors state that steerable antennas are of most use when ground stations are immediately below the HAPS node, with no benefit for stations that are on the edge of coverage.

From a macroscopic perspective, such intra- and inter-layer considerations, together with the misalignment challenges, would practically affect the true system performance metrics (e.g., coverage, rate, delay) and need to be addressed as part of the challenging cloud-enabled HAPS systems. One possible solution is to combine multiple spectrum enablers such as microwave, mmWave, terahertz (THz), and FSO communications. Each of these technologies has its advantages and disadvantages. On the one hand, lower-frequency microwave communications support longer communication distances and wider coverage, both of which are critical for the HAPS operation in the access domain. On the other hand, higher-frequency mmWave and THz communications support much higher data rates which are critical for the HAPS operation in the backhaul domain. Since extending the communication distance of mmWave/THz requires, given the high path and absorption losses, high antenna and beamforming gains, the corresponding beamwidths are usually narrow, reducing coverage and intensifying the misalignment effect. Nevertheless, in quasi-static HAPS deployments, such links can still be established. Note that the particular use of THz communications motivates the merging of communications, localization, and sensing applications for HAPS operations \cite{sarieddeen2019generation}. Furthermore, THz links can be used inside HAPS-deployed data centers \cite{faisal2019ultra}, where the environment can be perfectly controlled to remove molecular components that result in significant THz absorption losses. Such wireless links in data centers can significantly reduce the complications of wired connectivity within the HAPS, which is hard to maintain and incurs an extra weight factor that can not be neglected in flying platforms.

\subsection{QoS Requirements}
The most important requirement from the user’s perspective is the QoS. In general, the cloud customer tends to choose the service that better satisfies his/her specific QoS requirements. Many previous works outlined a variety of QoS metrics for cloud computing, including availability, reliability, response time, scalability, elasticity, and fairness \cite{ardagna2014quality, bardsiri2014qos}. In brief, availability is related to the overall percentage of time during which a service is available, reliability is related to several factors such as functioning adequately; producing correct outputs; and resisting failures, response time is the period between when a customer makes a request and a response is returned, scalability (sometimes called capacity) is related to the volume (or load) that a service can handle while maintaining standards of quality and performance, elasticity is the ability to increase/decrease the amount of service resources (such as CPU; storage; memory; and input/output bandwidth) on demand, and fairness is related to assigning correct priorities to customers and servicing them fairly based on their priorities in order to achieve overall customer satisfaction. On the other hand, several papers \cite{samht2015quality, alsamhi2016implementation} discussed the QoS from the HAPS perspective and illustrated the various methods that can enhance the QoS of the HAPS communications and deployment.

Most cloud platforms face challenges in meeting the various QoS requirements; the same is expected for the cloud services offered within the HAPS. First, a reliable connection should be maintained between the user’s device and the HAPS node to achieve the availability metric. This can be realized by installing HAPS gateways at high locations (such as the top of buildings) and establishing a high-speed connection between the users’ devices and the HAPS gateways. However, this cannot always be guaranteed, especially when the user consumes the service from a mobile device. When the user is mobile, the connection to the HAPS might be weak or lost in specific locations (such as underground tunnels), which decreases the availability of the HAPS cloud service. On the other hand, UAV base stations and TBs could be used as HAPS gateways in rural areas to increase the availability of HAPS cloud services in these areas.

Enhancing reliability is also a challenge with the advancement in the various cloud-related technologies. The work in \cite{gill2018failure} proposes a comprehensive taxonomy of reliability issues and a conceptual model for reliable cloud services. In addition to the mentioned issues, HAPS-based services face additional failure possibilities due to factors unique to the HAPS environment, such as weather conditions and unexpected accidents (such as objects colliding with the HAPS node). These factors should be considered, and appropriate solutions should be deployed to increase the reliability of the HAPS cloud services. As for the response time, it is mostly related to the communication techniques that will be used to connect the users’ devices with the HAPS. As stated before, intermediate nodes or networks (such as HAPS gateways in urban areas and UAVs and TBs in rural areas) could be used to enhance the connection between the user and the HAPS. Future studies should compare the various communication methods and their corresponding response time ranges to identify the suitable communication method for each type of cloud service according to its response time requirements. 

The scalability and elasticity of the HAPS cloud are expected to be limited as compared to ground data centers due to the restricted space within the HAPS for various cloud-required infrastructures (servers, network devices, storage resources, etc.). Hence, important factors to consider are the limits that should be specified for each service according to its computing and storage requirements. For example, the HAPS network contains, at a certain time, a specific number of nodes that accommodate a certain maximum amount of data. Data-intensive cloud applications that produce continuous streams of big data (such as real-time traffic management) should be configured to manage their storage requirements in order not to overwhelm the HAPS data center and affect the performance of other services. In general, cloud applications in the HAPS should be carefully designed and implemented to consider the limitation of the HAPS cloud in terms of scalability and elasticity.

Finally, fairness is one of the most difficult QoS metrics to satisfy due to the complex factors that affect it. Several studies \cite{ghodsi2011dominant, kohsuwan2013focal} discuss the elements that affect cloud service fairness and the methods that cloud providers should use to develop structural and social service fairness and guarantee equal service delivery and fair treatment. The same factors and requirements apply to the HAPS cloud. However, the limited amount of resources within the HAPS provides an additional challenge of allocating them fairly to the various applications and their users.

\subsection{Physical Maintenance}  
The cloud-enabled HAPS architecture proposed in this paper requires the deployment of several types of infrastructure (such as strong processors, data storage systems, and communication equipment) which would add to the HAPS platform  weight and consumed energy. Therefore, the maintenance of such systems would be an intricate mission, especially given its aerostatic standing within the stratospheric layer. While HAPS systems are, generally speaking, easier to bring back to to the ground level as compared to upper satellites structures \cite{kurt2020vision}, periodic maintenance of aerostatic HAPS nodes remains a necessity, especially in HAPS missions that require long endurance, which could be in the order of months or years \cite{stratobus}. 

Since the HAPS system envisioned in this paper necessitates powerful processing capabilities and connectivity premises, all while ensuring a prolonged, reliable, and effective system operation, the durable deployment of such system becomes a challenging function of the joint regular maintenance of the HAPS computing and communication infrastructure. This is especially true due to the QoS requirements that we mentioned in the previous section. Since many hazards could abrupt the HAPS normal operations (such as hardware failures, software bugs, and collisions), the continuous maintenance of the various HAPS infrastructure is vital for ensuring reliable connectivity, service availability, and acceptable response time of the HAPS cloud platform. All these factors should be taken into consideration in assessing practical network designs.

\subsection{Security Requirements} 
Since cyber threats are continuously growing and becoming more advanced, cloud customers deem it necessary to select a cloud platform that provides top-of-the-range security that suits the customer’s needs. In general, cloud security relates to the technologies, rules, disciplines, and services that safeguard the cloud software, framework, and data from security attacks. Researchers from academia, industry, and standardization societies have presented various perspectives on the cloud security necessities, identifying vulnerabilities, well-known threats, and potential responses and solutions \cite{singh2017cloud, kumar2019cloud, krutz2010cloud, mershad2011react}. The work in \cite{kumar2019cloud} emphasizes the fact that security should be enforced across all seven layers within the cloud network. The authors identify seven main requirements of cloud security: Confidentiality, integrity, availability, authentication, authorization, accountability, and privacy. The main threats that should be considered by cloud providers when designing a strong security framework are related to data breaches, account hijacking and identity theft, system and code vulnerabilities, malicious users, abuse utilization of services, and denial of service (DoS). Furthermore, the authors identify possible solutions to these threats, including encryption frameworks to manage identities, keys, and signatures, intrusion detection and prevention systems, software and virtual environment security, network security, storage security, physical and hardware security, and trust management systems. The various types of attacks that could be performed on the cloud framework, services, and users are identified in \cite{singh2017cloud}; they can be summarized as: DoS, service injection, user-to-root, virtualization, man-in-the-middle, phishing, metadata spoofing, and backdoor channel attacks. 

With the advancement of technology, new types of security systems are being researched and applied to the cloud, such as blockchain-based \cite{yang2020authprivacychain, park2017blockchain, medhane2020blockchain} and ML-based \cite{subramanian2019focus, kwabena2019mscryptonet, ma2018pdlm} models. According to \cite{yang2020authprivacychain}, the cloud provider can encrypt and store the cloud access control rights in the blockchain, which highly safeguards the privacy of users. In addition, when applying the blockchain to the cloud, the cloud transactions’ data are encrypted, distributed across the blockchain nodes, and interconnected via a hashing function \cite{park2017blockchain}. Hence, it becomes difficult for attackers to interfere with the transactions’ data due to the consensus mechanism of the blockchain. The authors of \cite{subramanian2019focus} propose using a convolution neural network (CNN) to analyze the cloud traffic data, detect and identify sensitive data patterns, and determine for each new data flow whether it belongs to normal or abnormal activities. Furthermore, a collaborative privacy-preserving ML architecture for healthcare systems is presented in \cite{kwabena2019mscryptonet}. The proposed mechanism provides important solutions to several challenges that exist in ML-based ubiquitous healthcare applications. Applying deep learning over cloud data encrypted under multiple keys is further proposed in \cite{ma2018pdlm}. In this approach, the cloud provider migrates the users’ data that are encrypted with the users’ keys to the deep learning training model, and the latter trains the model using the stochastic gradient descent (SGD) algorithm and performs the feed-forward and back-propagation procedures based on a privacy-preserving calculation toolkit. 

Since the cloud-enabled HAPS faces all these security challenges and threats, it is important to study the best security mechanisms that should be applied within the HAPS network to secure the various infrastructures, applications, and data on each HAPS node. In addition, the connections within the HAPS network and between the HAPS and the various entities (Fig. \ref{FIG:allServices}) should be secured against network-related attacks. In general, it is expected that a collection of security measures and procedures will be required to secure the various elements of the HAPS cloud system. The selection of the most appropriate, efficient, and secure mechanisms for each element of the HAPS cloud system is expected to become a hot research topic in the near future.

\section{Future Research Directions}
\label{Sec:future}

\begin{table*}[htp]
\caption{Summary of the proposed future research directions.}
\label{Tab:FutureSummary}
\centering
\begin{tabular}{p{0.14\linewidth}p{0.38\linewidth}p{0.38\linewidth}}
\hline
\textbf{Future Research Direction}	& \textbf{Key Points} &	\textbf{Objectives} \\
\hline
HAPS Mobility and Message Routing	&
\begin{itemize}[leftmargin=*,noitemsep,partopsep=0pt,topsep=0pt,parsep=0pt]
    \item A framework for routing packets within the HAPS network and between the HAPS and the surrounding networks 
    \item An optimization problem that takes as inputs the locations and movements of nodes and the environmental conditions and produces the required locations of the HAPS nodes
\end{itemize}
&
\begin{itemize}[leftmargin=*,noitemsep,partopsep=0pt,topsep=0pt,parsep=0pt]
    \item Managing the intra-HAPS and inter-HAPS connections
    \item Establishing routing paths between the HAPS nodes
    \item Controlling the mobility of the HAPS nodes in order to maintain the connections with the surrounding networks 
\end{itemize}
 \\
\hline
HAPS Security via Blockchain and Machine Learning
&
\begin{itemize}[leftmargin=*,noitemsep,partopsep=0pt,topsep=0pt,parsep=0pt]
    \item A blockchain model for securely saving the data and transactions within the HAPS cloud
    \item A consensus mechanism suitable for saving the energy of the HAPS nodes
    \item A ML model for monitoring the blockchain operations	
\end{itemize}
&
\begin{itemize}[leftmargin=*,noitemsep,partopsep=0pt,topsep=0pt,parsep=0pt]
    \item Establishing a distributed ledger for saving financial transactions between the HAPS-cloud customers and providers 
    \item Establishing a data blockchain for securely saving all data operations that occur within the HAPS cloud platform
    \item Detecting and resolving security attacks that target the blockchain communications and data
\end{itemize}
 \\
\hline
Intelligent Cloud-Enabled HAPS
&
\begin{itemize}[leftmargin=*,noitemsep,partopsep=0pt,topsep=0pt,parsep=0pt]
    \item A system that combines edge computing and ML capabilities 
    \item Network slicing is used to create multiple virtual instances within the HAPS cloud, and optimize each instance based on the requirements of each specific service
\end{itemize}
&
\begin{itemize}[leftmargin=*,noitemsep,partopsep=0pt,topsep=0pt,parsep=0pt]
    \item Making the HAPS cloud dynamically adapt to various network and services changes
    \item Identifying and implementing the ML algorithm that is most suitable to monitor the performance of each HAPS cloud service and adjust its execution parameters
\end{itemize}
 \\
\hline
Artificial Intelligence-Based Resource Allocation
&
\begin{itemize}[leftmargin=*,noitemsep,partopsep=0pt,topsep=0pt,parsep=0pt]
    \item A framework that utilizes AI techniques to allocate various HAPS resources correctly and efficiently
\end{itemize}
&
\begin{itemize}[leftmargin=*,noitemsep,partopsep=0pt,topsep=0pt,parsep=0pt]
    \item Achieving good trade-offs between the HAPS-cloud different functionalities, such as data rate, power, energy, latency, coverage, and security
\end{itemize}
 \\
\hline
Integration with Vertical Heterogeneous Networks
&
\begin{itemize}[leftmargin=*,noitemsep,partopsep=0pt,topsep=0pt,parsep=0pt]
    \item A model to manage the real-time cloud-related operations and communications between the integrated space-air-ground networks
\end{itemize}
&
\begin{itemize}[leftmargin=*,noitemsep,partopsep=0pt,topsep=0pt,parsep=0pt]
    \item Creating a multi-cloud system that harvests the advantages of each of the space-air-ground networks and combines their cloud platforms into a distributed super cloud
\end{itemize}
 \\
\hline
\end{tabular}
\end{table*}

In light of the challenges described above, many fundamental issues need to be addressed so as to fully assess the gains harvested from the practical deployment of cloud-enabled HAPS. This section presents the main open issues related to the design of such systems, namely, HAPS mobility and message routing, HAPS security via blockchain and ML, AI-based resource allocation in cloud-enabled HAPS, and integration with vertical heterogeneous networks.

\subsection{HAPS Mobility and Message Routing}
\label{Sec:future_Mobility_And_Routing}
The deployment of HAPS will not succeed without a solid framework that identifies and applies the correct rules and procedures for the mobility of the HAPS nodes. Such framework should optimize the routing of packets within the HAPS network (intra-HAPS communications) and between the HAPS nodes and other elements in Fig.~\ref{FIG:allServices} (inter-HAPS communications). Hence, we aim to design and implement a routing protocol that calculates the optimal routing path from a source to a destination HAPS node based on the routing requirement. The latter could be the minimum delay (which is achieved by selecting the shortest path that contains the minimum average distance between adjacent HAPS nodes), the highest QoS (for instance, delivering the data with high accuracy), the least energy consumption (for low priority messages), etc. In addition, the routing algorithm monitors the connections between the HAPS nodes and the surrounding networks and suggests to the HAPS control center adjusting the location of one or more HAPS nodes in order to maintain or enhance the various connections between the HAPS network and the satellite, UAV, TBs, and ground networks. Since some of these latter networks are mobile, the HAPS connection to one or more external networks could be lost under mobility. Hence, it is essential to move some HAPS nodes to new locations to maintain the connections with these networks.

In order to achieve these objectives, an optimization problem should be solved to calculate the best position of each HAPS node that satisfies the intra-HAPS routing requirements and ensures the persistence of the inter-HAPS connections. The optimization problem is re-executed with updated parameters whenever the status and environment of one of these networks changes (satellites’ movements, UAV nodes’ movements, new topographic conditions, environmental changes, etc.). In all these cases, the optimization problem will be re-calculated with the new parameters that correspond to the new locations of various nodes and/or the new connectivity conditions (such as lost links and new links). The optimization problem results will be the locations of the HAPS nodes and the new intra- and inter-routing paths. Hence, these results could require moving one or more HAPS nodes to new locations and establishing new routing paths. The optimization problem should take into consideration the limits on the HAPS node movement that the HAPS network administrators predefine.

\subsection{HAPS Security via Blockchain and Machine Learning}
\label{future_Security}
We propose a future research direction in which we apply a blockchain model augmented with ML to secure the data and access the HAPS cloud services. The model utilizes blockchain capabilities to encrypt/decrypt all the data exchanged between cloud users and providers. In particular, the model can include two blockchains: The first is a distributed ledger that securely saves all financial transactions in the cloud. Each financial transaction is encrypted by the public key of the receiving HAPS node and added to the next block. The HAPS nodes constitute the blockchain miners that generate the blockchain blocks. In order to do that, the HAPS nodes will execute a consensus model in which one of the HAPS nodes is selected as the current miner that generates the next block. The role of the current miner alternates between the HAPS nodes. Each HAPS node validates each financial transaction it receives from the customers and sends it to the current miner. A new block is generated each period \textit{T\textsubscript{block}}. When the current miner generates the new block, it broadcasts it to the HAPS nodes and waits for their acknowledgments. Each HAPS node checks the validity of the transactions in the new block by validating the signature of the cloud customer(s) who made the transaction. If all transactions are valid, the HAPS node sends an acknowledgment to the current miner. When the current miner receives acknowledgments from \textit{TH\textsubscript{C}} nodes (where \textit{TH\textsubscript{C}} is a threshold determined by the consensus algorithm), it considers the new block valid and broadcasts its ID to the HAPS network so that all nodes add it to the blockchain. In this way, all financial transactions within the HAPS cloud services are saved securely in the cloud ledger. A similar model for the IoV was presented in \cite{mershad2021proof}.

The other blockchain in our model is the data blockchain. Here, it is not feasible to save all data exchanged between cloud providers and users in the blockchain since some applications generate huge amounts of data and some users store extensive volumes of data in the cloud. Instead, the data in each cloud transaction is hashed, and the hash result is saved in the data blockchain as the transaction ID. In addition, this ID will be linked to the actual data that is encrypted and saved in the cloud distributed database. The data could be hashed hierarchically in multiple iterations if it is too big. Hence, the hash of each data transaction will be saved in the data blockchain for future reference. Whenever the cloud provider or the user wants to review the data operations of a previous time period, the hashes can be fetched from the blockchain and used to index the distributed database and obtain the data of the corresponding transactions. This operation is required whenever there is a need to check the details of the data transactions or to resolve any dispute between the user and the cloud provider.

While the blockchain is used to save the data and financial transactions of the cloud securely, a ML model can be used to manage and monitor the blockchain operations to detect any attack attempts on the blockchain data and messages. For example, suppose an attacker attempts to insert malicious transactions in the next block by compromising multiple nodes in the blockchain. In that case, the ML algorithm should be able to detect the attack from the behavior of the nodes in the blockchain network. The ML algorithm will continuously monitor the exchanged messages and their encrypted content and compare them to attacks’ signatures and behaviors in the algorithm’s training data. In general, the ML model can be used to monitor all the operations on the blockchain network to detect new attacks, depending on a set of training data that contains attack samples. When new attacks are detected, they are added to the training data to detect similar future attacks.

\subsection{Intelligent Cloud-Enabled HAPS}
Edge computing is the process of shifting part of the cloud computations near the data source or location. In general, edge computing services reduce data bandwidth needs, improve latency, and ease cloud computing requirements. By supporting Network Function Virtualization (NFV), network functions can be divorced from dedicated hardware, allowing Software Defined Networking (SDN) to rapidly deploy services as needed. NFV/SDN and edge computing share a symbiotic relation where, mutually, they create a framework for modularized, re-configurable network adaptability that can easily support network slicing \cite{taleb2017multi, foukas2017network}. Using network slicing, multiple network instances can be created, each optimized for a specific service, yet sharing the same physical infrastructure. While these innovations create massive adaptability, they also create an exponentially expanding design space that is difficult to observe, control, and predict, which is exactly where innovative Artificial Intelligence and Machine learning (AI/ML) techniques are expected to make the most contributions. AI/ML are expected to fuel the next phase of innovation at the cloud-enabled HAPS systems where both the wireless networks and infrastructure become more dynamic, virtualized, intelligent, distributed, and energy-efficient.

In this context, ML is exceptionally good at solving problems where the design space is extremely large, and complicated rule lists can be replaced by elegant ML routines that can learn from previous data using automated feature extraction to alleviate the need for tedious manual extraction. In such high-altitude cloud systems, not only is the parameter space huge, it is also time-varying, and more often than not, involves objective functions that are non-linear and/or polynomial, requiring complicated heuristics based on expensive measurement campaigns. ML offers a solution for setting and continually updating network parameters based on actual usage patterns. Generally speaking, Edge Machine Learning (EML) can be used for training, inference, and Mobile Multi-Access Edge computing (MEC), all aimed at providing lower latency, higher reliability, and distributed network resources. However, EML faces significant challenges, including long training times, high computational power requirements, difficulty to generate and process sufficient relevant training samples, scalability, among other issues that have limited adoption.

There is no ML technique that can be said to be better than any other. According to the No Free Lunch theorem, any two methodologies may perform equally well in solving a problem if there is no prior information about the problem. However, the selection of certain learning techniques for optimization in cloud-enabled HAPS is tied with the services, the ultimate objective of that service, and the application information domain. For example, artificial neural networks (ANNs) fit predicting service while reinforcement learning is more suitable for optimization with a certain goal such as service latency. Reinforcement learning matches services in absence of a priori knowledge and a sufficiently large amount of training data. However, they suffer from a major drawback: high calculation cost because an optimal solution requires that all states be visited to choose the optimal one. Thus, this tradeoff must be carefully considered when implementing reinforcement learning in certain environments. As shown in \cite{witten2011data}, support vector machine (SVM) based learning outperforms deep learning in terms of computation time for training in the experiment of predicting the next location of a user, considering the associated delays. However, deep learning wins by far in predicting the number of users located in a certain location. Deep learning requires 27 mins to train while SVM consumed 6 days for training \cite{witten2011data}. Thus, we can conclude that each ML technique has advantages and drawbacks that make it performs better in one application but not in the other.

The selection of ML platforms to be utilized is a challenging task as there is no single platform that can fit all solutions. Each platform has advantages and drawbacks. Thus, there are important factors to be considered in the selection of a ML platform, which are scalability, speed, coverage, usability, extensibility, and programming languages support. The size and complexity of data are the main drivers of the scalability factor. The processing platform that the library is running impacts the speed factor. Coverage comprises the number of ML algorithms implemented in the tool. The usability factor includes initial setup, maintenance, and the availability of programming language. The extendibility factor means that the implementation used in the toolkit can be useful in the design of a new platform. However, some application requires real-time data for analysis. Thus, stream processing is designed to analyze and act on real-time streaming data, using continuous queries to handle high volume in real-time with a scalable, highly available, and fault-tolerant architecture. With all these available ML platforms, an important design decision should be made considering the application requirements and the associated constraints of the cloud-enabled HAPS architecture.

\subsection{Artificial Intelligence-Based Resource Allocation}
The proposed cloud-enabled HAPS system performance is a strong function of the practicality of interference management techniques, especially developed to manage the network radio resources. Such techniques, which are expected to be coordinated across the HAPS, air, and ground networks, aim to achieve good trade-offs between the network's different functionalities, e.g., rate, energy efficiency, power, latency, coverage, and security. Such utilities are all functions of how to best design the system parameters, e.g., the spectrum utilized by the different system links, the transmit power at the HAPS layer, the traffic routing protocols, the inter-HAPS links, as well as the edge and core processing capabilities of the system under study, which makes the resource allocation problem in cloud-enabled HAPS intricate in nature. Given the additional dynamics of the ground communications infrastructures and the stochasticity of the HAPS-to-ground and HAPS-to-air channels, we expect AI-based techniques to play a major role in the design and optimization of future cloud-enabled HAPS networks, as also mentioned in the previous section. The promises of ML in solving complex and analytically intractable optimization problems (e.g., see \cite{DahroujAccessML2021} and references therein) would indeed fuel a plethora of open issues in the context of dynamic resource optimization in cloud-enabled HAPS, which promises to be an active area for future research.

\subsection{Integration with Vertical Heterogeneous Networks}
As discussed earlier, cloud-enabled HAPS networks are expected to be a part of multi-layered integrated space-air-ground networks (VHetNets). Such integration includes geo-satellites, cubesats, HAPS, hot-air balloons, LAPS, and UAVs to best serve ground-level communications infrastructures. Such system performance becomes a strong function of the interactions between its different entities, their mutual dynamics, and the ever-changing environmental conditions of all the inter- and intra-connecting links, especially given the spatio-temporal variations of such distributed entities and their relative positioning. An important future research direction in this field is, therefore, to enable the real-time operations of such integrated space-air-ground networks, by jointly accounting for both the cloud computing capabilities of HAPS, and the distributed intelligence of other networks entities through invoking their wireless edge computing functionalities. For instance, such integration would bring other levels of computing into the overall integrated paradigm, e.g., fog-computing at the UAV level, and edge-computing at the network edge \cite{KanekoFRAN2020}. 

The coordination among such multi-layered flying networks would then require a tactful joint operational treatment which includes, but is not limited to, joint routing protocols and synchronization operations on the networking layers, joint resource allocation on the medium access control (MAC) layers, and joint modulation and coding schemes (MCS) designs on the physical layers. Such a compound infrastructure, although intricate,  would open the way for a multi-cloud system that harvests the advantages of each of the aforementioned networks, and combines their cloud platforms into a distributed super cloud that reaches a wider range of users, and offers them richer QoS services, which promises to be an exciting area of future research.\\

Finally, for the sake of completeness, we refer the readers to Table \ref{Tab:FutureSummary}, which summarizes the future research directions discussed in this section. More specfically, Table \ref{Tab:FutureSummary} illustrates the key points and objectives of each of the proposed research directions above.

\section{Conclusion}
\label{Sec:conclusion}
Augmenting ground-level communications with HAPS is expected to be among the major physical-layer innovations of 6G wireless systems. Given the HAPS quasi-static positioning at the stratosphere, HAPS-to-ground and HAPS-to-air connectivity frameworks are expected to be prolific in terms of data acquisition and computing, which motivates to equip HAPS with cloud computing capabilities. This paper first advocates for the potential physical advantages of deploying HAPS as flying data centers, also known as super-macro base stations. The paper then presents the merits that can be achieved by integrating various cloud services within the HAPS, and the corresponding cloud-type applications that would utilize the HAPS for enhancing the quality, range, and types of offered services. The paper further presents some of the challenges and resource allocation problems that need to be addressed for realizing practical cloud-enabled HAPS, mainly those related to the high energy and processing power, maintenance, QoS, and security considerations. Finally, the paper sheds light on some open research directions in the field, namely, HAPS mobility and message routing, HAPS security via blockchain and ML, AI-based resource allocation in cloud-enabled HAPS, and integration with vertical heterogeneous networks. To the best of the authors' knowledge, this paper is the first of its kind which advocates for the use of cloud-enabled HAPS from a holistic perspective, and presents the challenges and open issues of such deployment, which promises to be an active area of research in the context of 6G systems development and cross-layer optimization. 

\section*{Acknowledgment}
This work was supported in part by the Center of Excellence for NEOM Research at the King Abdullah University of Science and Technology (KAUST).

\ifCLASSOPTIONcaptionsoff
  \newpage
\fi



%



\bibliographystyle{IEEEtranN}
\bibliography{ref}

%






\end{document}